\documentclass[superscriptaddress,aps,preprint,nofootinbib,showpacs]{revtex4}
\usepackage{graphicx,color,amsmath, epsfig, rotating}
\newcommand{\be}{\begin{equation}}
\newcommand{\ee}{\end{equation}}
\newcommand{\ba}{\begin{eqnarray}}
\newcommand{\beq}{\begin{equation}}
\newcommand{\eeq}{\end{equation}}
\newcommand{\ea}{\end{eqnarray}}

\newcommand{\MNS}{{\text{MNS}}}
\newcommand{\eV}{\text{eV}}
\newcommand{\GeV}{\text{GeV}}
\newcommand{\TeV}{\text{TeV}}
\newcommand{\Hlilj}{H^{\pm\pm}\to l_i^\pm l_j^\pm}

\newcommand{\BR}{\text{BR}}

\newcommand{\magic}{{\text{mgc}}}

\newcommand{\tlll}{{\tau \to \bar{l}_i l_j l_k}}
\newcommand{\tmmm}{{\tau \to \bar{\mu}\mu\mu}}
\newcommand{\tmme}{{\tau \to \bar{\mu}\mu e}}
\newcommand{\temm}{{\tau \to \bar{e}\mu\mu}}
\newcommand{\tmee}{{\tau \to \bar{\mu}ee}}
\newcommand{\teme}{{\tau \to \bar{e}\mu e}}
\newcommand{\meee}{{\mu \to \bar{e}ee}}
\newcommand{\meg}{{\mu \to e\gamma}}
\newcommand{\ratio}[2]{\text{BR($#1$)/BR($#2$)}}
\def\beqa{\begin{eqnarray}}
\def\eeqa{\end{eqnarray}}

\def\bea{\begin{eqnarray}}
\def\eea{\end{eqnarray}}

\def\err#1#2{\lower2pt\hbox{ $\stackrel{\scriptstyle +#1}{\scriptstyle -#2}$}}
\def\ga{\mathrel{\raise.3ex\hbox{$>$\kern-.75em\lower1ex\hbox{$\sim$}}}}
\def\la{\mathrel{\raise.3ex\hbox{$<$\kern-.75em\lower1ex\hbox{$\sim$}}}}
\def\bmaT{\left(\begin{array}{ccc}}
\def\emaT{\end{array}\right)}
\def\bma{\left( \begin{array} }
\def\ema{\end{array} \right)}
\def\gsim{~{\rlap{\lower 3.5pt\hbox{$\mathchar\sim$}}\raise 1pt\hbox{$>$}}\,}
\def\lsim{~{\rlap{\lower 3.5pt\hbox{$\mathchar\sim$}}\raise 1pt\hbox{$<$}}\,}

\begin{document}

\preprint{TU-844}
\preprint{SISSA 21/2009/EP}

\title{\boldmath Lepton Flavour Violating Decays $\tau\to \bar{l}ll$ and $\mu\to e\gamma$ \\ in the Higgs Triplet Model  \unboldmath} 

\author{A.G. Akeroyd}
\email{akeroyd@ncu.edu.tw}
\affiliation{Department of Physics, National Central University, Jhongli 320, Taiwan}
\author{Mayumi Aoki}
\email{mayumi@tuhep.phys.tohoku.ac.jp}
\affiliation{Department of Physics, Tohoku University, Sendai 980-8578, Japan}
\author{Hiroaki Sugiyama}
\email{hiroaki@fc.ritsumei.ac.jp}
\affiliation{SISSA, via Beirut 2-4, I-34014 Trieste, Italy}
\affiliation{Department of Physics,
Ritsumeikan University, Kusatsu, Shiga 525-8577, Japan}


\begin{abstract}
Singly and doubly charged Higgs bosons in the Higgs Triplet Model
mediate the lepton flavour violating (LFV) decays 
$\tau\to \bar{l}ll$ and $\mu\to e\gamma$. The LFV decay rates 
are proportional to products of two triplet Yukawa couplings ($h_{ij}$) which can be expressed
in terms of the parameters of the neutrino mass matrix and an unknown triplet
vacuum expectation value. We determine the parameter space of the
neutrino mass matrix in which a signal for $\tau\to \bar{l}ll$ and/or 
$\mu\to e\gamma$ is possible at ongoing and planned experiments.
The conditions for  respecting the stringent upper limit for 
$\meee$ are studied in detail, with emphasis given to the possibility 
of $|h_{ee}|\simeq 0$
which can only be realized if Majorana phases are present.
\end{abstract}

\maketitle


\section{Introduction} 
The now-established evidence that neutrinos oscillate and possess a small mass
below the eV scale~\cite{Fukuda:1998mi}
necessitates physics beyond the Standard Model
(SM), which could 
manifest itself at the CERN Large Hadron Collider (LHC) 
and/or in low energy experiments which 
search for lepton flavour violation (LFV)~\cite{Kuno:1999jp}.
Consequently, models of neutrino mass generation which can be probed
at present and forthcoming experiments are of great phenomenological
interest. 

Neutrinos may obtain mass via the vacuum expectation value
(vev) of a neutral Higgs boson in an isospin triplet
representation~\cite{Konetschny:1977bn, Mohapatra:1979ia, Magg:1980ut,Schechter:1980gr,Cheng:1980qt}.
A particularly simple implementation of this
mechanism of neutrino mass generation is the 
``Higgs Triplet Model'' (HTM)
in which the SM Lagrangian is augmented solely by
an $SU(2)$ triplet of scalar particles 
with hypercharge $Y=2$~\cite{Konetschny:1977bn, Schechter:1980gr,Cheng:1980qt}.
In the HTM neutrinos acquire a Majorana mass
given by the product of a triplet Yukawa coupling ($h_{ij}$)
and a triplet vev ($v_\Delta$).
Consequently, there is a direct connection 
between $h_{ij}$ and the neutrino mass matrix which gives rise to 
phenomenological predictions for processes which depend on $h_{ij}$.
A distinctive signal of the HTM would be the observation of doubly charged Higgs bosons ($H^{\pm\pm}$)
whose mass ($m_{H^{\pm\pm}}$) may be of the order of the electroweak scale. Such particles 
can be produced with sizeable rates at hadron colliders in the
processes $q\overline q\to H^{++}H^{--}$ \cite{Gunion:1989in}
and $qq'\to H^{\pm\pm}H^{\mp}$~\cite{Dion:1998pw, Akeroyd:2005gt}.
Direct searches have been carried out at the Fermilab Tevatron in the 
production channel
$q\overline q\to H^{++}H^{--}$ and decay $H^{\pm\pm}\to l^\pm_il^\pm_j$, 
with mass limits of the order $m_{H^{\pm\pm}}> 110\to 150$~GeV~\cite{Acosta:2004uj}.
The branching ratios (BRs) for 
$H^{\pm\pm}\to l^\pm_il^\pm_j$ depend on $h_{ij}$
and are predicted in the HTM in terms of the parameters
of the neutrino mass matrix%
~\cite{Akeroyd:2005gt, Ma:2000wp, Chun:2003ej}.
Detailed quantitative studies of
BR($H^{\pm\pm}\to l^\pm_il^\pm_j$) in the HTM have been performed in
\cite{Garayoa:2007fw,Akeroyd:2007zv,Kadastik:2007yd,Perez:2008ha} 
with particular emphasis  
given to their sensitivity to the Majorana phases and 
the absolute neutrino mass, i.e., parameters which cannot be 
probed in neutrino oscillation experiments. Recent simulations \cite{Perez:2008ha, delAguila:2008cj}
of the discovery prospects of $H^{\pm\pm}$ at the LHC now include
the mechanism $qq'\to H^{\pm\pm}H^{\mp}$, 
which plays a crucial role in extracting the parameters of the 
neutrino mass matrix.

The Yukawa couplings $h_{ij}$ also mediate low energy 
lepton flavour violating (LFV) processes. 
In this paper we study the BRs of the LFV decays
$\tlll$ and $\meee$ (which are mediated 
by $H^{\pm\pm}$) and $\mu\to e\gamma$ (which is mediated by 
$H^{\pm\pm}$ and $H^\pm$). Previous studies 
of such decays in the HTM were performed in 
\cite{Chun:2003ej, Kakizaki:2003jk}, and it was shown that
specific patterns of LFV are predicted in analogy with 
the prediction for BR$(H^{\pm\pm}\to l^\pm l^\pm)$.
Experimental prospects for $\mu\to e\gamma$ are bright with the
recent commencement of the MEG experiment which will
probe BR$\sim 10^{-13}$, two orders of magnitude beyond
the current upper limit~\cite{Grassi:2005ac}.
 The decay $\tlll$ is
currently being searched for at the $e^+e^-$ $B$ factories  
with upper limits in the range
BR($\tlll) <  2 \to 8\times 
10^{-8}$~\cite{Abe:2007ev, Aubert:2007pw}. 
Simulations of the detection prospects
at a proposed high luminosity $e^+e^-$ $B$ factory with 
${\cal L} = 5 \to 75$~ab$^{-1}$ anticipate sensitivity to
BR$\sim 10^{-9}\,\text{-}\,10^{-10}$~\cite{Hashimoto:2004sm,Bona:2007qt,
Browder:2007gg,Browder:2008em}.
Searches for $\tmmm$ can be performed at
the LHC where $\tau$ leptons are copiously produced
from the decays of $W,Z,B,D$, with anticipated sensitivities to 
BR$\sim 10^{-8}$~\cite{Santinelli:2002ea,Giffels:2008ar}.

The decay $\meee$, for which there is
a strict bound BR$< 10^{-12}$ \cite{Bellgardt:1987du}, is a strong
constraint on the parameter space of $h_{ij}$ in the HTM\@. 
Obtaining BR($\tlll)> 10^{-9}$ together
with compliance of the above bound on BR($\meee$)
is possible but can only be realized in specific regions of the parameter
space for $h_{ij}$ (e.g., $|h_{e\mu}|\simeq 0$ \cite{Chun:2003ej}).
In this paper we 
perform a detailed quantitative study in order to find the
parameter space of the neutrino mass matrix where 
a signal for $\tlll$ and/or $\mu\to e\gamma$ is possible at ongoing 
and planned experiments.  We study in detail a novel way to satisfy the
constraint from  $\meee$, namely $|h_{ee}|\simeq 0$, which can only be realized
if Majorana phases are present \cite{Akeroyd:2006bb}. 
The pattern of LFV violation for $|h_{ee}|\simeq 0$ is 
studied and shown to differ from that for the case of $|h_{e\mu}|\simeq 0$. We also
discuss the prediction for BR($H^{\pm\pm}\to l^\pm_il^\pm_j$) in the HTM 
if a signal for $\tlll$ and/or $\mu\to e\gamma$ is observed.

Our work is organized as follows. In section II the
HTM is briefly reviewed. In section III the theoretical
basis for the decays BR$(\tlll)$ and BR($\mu\to e\gamma$) is
presented. The numerical analysis is contained 
in section IV with conclusions given in section V\@.

\section{The Higgs Triplet Model}

In the Higgs Triplet Model~(HTM)~\cite{Konetschny:1977bn,Schechter:1980gr,Cheng:1980qt}
a $I=1,Y=2$ complex $SU(2)_L$ isospin triplet of 
scalar fields is added to the SM Lagrangian. 
Such a model can provide a Majorana mass for the observed neutrinos 
without the introduction of a right-handed neutrino via the 
gauge invariant Yukawa interaction:
\begin{equation}
{\cal L}=h_{ij}\psi_{iL}^TCi\tau_2\Delta\psi_{jL}+h.c
\label{trip_yuk}
\end{equation}
Here $h_{ij} (i,j=e,\mu,\tau)$ is a complex
and symmetric coupling,
$C$ is the Dirac charge conjugation operator, $\tau_2$
is a Pauli matrix,
$\psi_{iL}=(\nu_i, l_i)_L^T$ is a left-handed lepton doublet,
and $\Delta$ is a $2\times 2$ representation of the $Y=2$
complex triplet fields:
\begin{equation}
\Delta
=\bma{cc}
\Delta^+/\sqrt{2}  & \Delta^{++} \\
\Delta^0       & -\Delta^+/\sqrt{2}
\ema
\end{equation}
A non-zero triplet vacuum expectation value $\langle\Delta^0\rangle$ 
gives rise to the following mass matrix for neutrinos:
\begin{equation}
m_{ij}=2h_{ij}\langle\Delta^0\rangle = \sqrt{2}h_{ij}v_{\Delta}
\label{nu_mass}
\end{equation}
The necessary non-zero $v_{\Delta}$ arises from the minimization of
the most general $SU(2)\otimes U(1)_Y$ invariant Higgs potential,
which is written as follows~\cite{Ma:2000wp, Chun:2003ej}
(with $\Phi=(\phi^+,\phi^0)^T$):
\begin{eqnarray}
V&=&m^2(\Phi^\dagger\Phi)+\lambda_1(\Phi^\dagger\Phi)^2+M^2
{\rm Tr}(\Delta^\dagger\Delta) +
\lambda_2[{\rm Tr}(\Delta^\dagger\Delta)]^2+ \lambda_3{\rm Det}
(\Delta^\dagger\Delta)  \nonumber \\
&&+\lambda_4(\Phi^\dagger\Phi){\rm Tr}(\Delta^\dagger\Delta)
+\lambda_5(\Phi^\dagger\tau_i\Phi){\rm Tr}(\Delta^\dagger\tau_i
\Delta)+\left(
{1\over \sqrt 2}\mu(\Phi^Ti\tau_2\Delta^\dagger\Phi) + h.c \right)
\label{higgs_potential}
\end{eqnarray}
Here $m^2<0$ in order to ensure $\langle\phi^0\rangle=v/\sqrt 2$, which
spontaneously breaks $SU(2)\otimes U(1)_Y$
to  $U(1)_Q$, and $M^2\,(>0)$ is the mass term for the triplet scalars.
In the model of Gelmini-Roncadelli~\cite{Gelmini:1980re} 
the term $\mu(\Phi^Ti\tau_2\Delta^\dagger\Phi)$ is absent,
which leads to spontaneous violation of lepton number for $M^2<0$.
The resulting Higgs
spectrum contains a massless triplet scalar (majoron, $J$) and another light 
scalar ($H^0$). Pair production via $e^+e^-\to H^0J$ would give a large 
contribution to the invisible width of the $Z$ and this model
was excluded at the CERN Large Electron Positron Collider (LEP). 
The inclusion of the term $\mu(\Phi^Ti\tau_2\Delta^\dagger\Phi$)
explicitly breaks lepton number when $\Delta$ is assigned $L=2$, and eliminates
the Majoron~\cite{Konetschny:1977bn,Schechter:1980gr,Cheng:1980qt}.
 Thus the scalar potential in
eq.~(\ref{higgs_potential}) together with the triplet Yukawa interaction of
eq.~(\ref{trip_yuk}) lead to a phenomenologically viable model of neutrino mass
generation.
 For small $v_\Delta/v$,
the expression for $v_\Delta$
resulting from the minimization of $V$ is:
\begin{equation}
v_\Delta \simeq \frac{\mu v^2}{2M^2+(\lambda_4+\lambda_5)v^2} \ .
\label{tripletvev}
\end{equation}
For large $M$ compared to $v$,
one has $v_\Delta \simeq \mu v^2/2M^2$
which is sometimes referred to as the ``Type II seesaw mechanism''
and would naturally lead to a small $v_\Delta$.
Recently there has been much interest in 
the scenario of light triplet scalars ($M\approx v$) 
within the discovery reach of
the LHC, for which eq.~(\ref{tripletvev}) leads to $v_\Delta\approx \mu$.
In extensions of the HTM the term $\mu(\Phi^Ti\tau_2\Delta^\dagger\Phi$) 
may arise in various ways: i) it could be from the vev of a Higgs singlet field%
~\cite{Schechter:1981cv}; 
ii) it could be generated at higher orders in 
perturbation theory~\cite{Chun:2003ej};
iii) it could originate in the context of extra dimensions \cite{Ma:2000wp}.

An upper limit on $v_\Delta$ can be obtained from
considering its effect on the parameter $\rho (=M^2_W/M_Z^2\cos^2\theta_W)$. 
In the SM $\rho=1$ at tree-level, while in the HTM one has
(where $x=v_\Delta/v$):
\begin{equation}
\rho\equiv 1+\delta\rho={1+2x^2\over 1+4x^2}
\label{deltarho}
\end{equation}
The measurement $\rho\approx 1$ leads to the bound
$v_\Delta/v\lsim 0.03$, or  $v_\Delta<8$~GeV\@.
At the 1-loop level $v_\Delta$ must be renormalized, and explicit
analyses lead to bounds on its magnitude similar to those derived from
the tree-level analysis, e.g.\ see \cite{Blank:1997qa}.
The HTM has seven Higgs bosons $(H^{++},H^{--},H^+,H^-,H^0,A^0,h^0)$.
The doubly charged $H^{\pm\pm}$ is entirely composed of the triplet 
scalar field $\Delta^{\pm\pm}$, 
while the remaining eigenstates are, in general, mixtures of the  
doublet and triplet fields. Such mixing is proportional to the 
triplet vev, and hence small {\it even if} $v_\Delta$
assumes its largest value of a few GeV\@.
Therefore $H^\pm,H^0,A^0$ are predominantly composed 
of the triplet fields, while $h^0$ is predominantly composed of the 
doublet field and plays the role of the SM Higgs boson.
The squared masses of $H^{\pm\pm},H^\pm,H^0,A^0$ are of order $M^2$ with splittings 
of order $\lambda_5 v^2$. The mass hierarchy $m_{H^{\pm\pm}} < m_{H^\pm} < 
m_{H^0,A^0}$ is obtained for $\lambda_5 > 0$
($m_{H^{\pm\pm}} > m_{H^\pm} > m_{H^0,A^0}$ for $\lambda_5 < 0$).

The phenomenologically attractive feature of the HTM is the direct connection
between the triplet Yukawa coupling $h_{ij}$ and the neutrino mass matrix ($m_{ij}$)
shown in eq.~(\ref{nu_mass}). The mass matrix for three Dirac neutrinos
is diagonalized by the MNS (Maki-Nakagawa-Sakata) matrix
$V_\MNS$~\cite{Maki:1962mu}
for which the standard parametrization is:
\begin{equation}
V_\MNS^{} =
\bmaT
c_{12}c_{13}                        & s_{12}c_{13}                  & s_{13}e^{-i\delta} \\
-s_{12}c_{23}-c_{12}s_{23}s_{13}e^{i\delta}  & c_{12}c_{23}-s_{12}s_{23}s_{13}e^{i\delta}  & s_{23}c_{13} \\
s_{12}s_{23}-c_{12}c_{23}s_{13}e^{i\delta}   & -c_{12}s_{23}-s_{12}c_{23}s_{13}e^{i\delta} & c_{23}c_{13}  
\emaT
\,,
\end{equation} 
where $s_{ij}\equiv\sin\theta_{ij}$ and $c_{ij}\equiv \cos\theta_{ij}$,
and $\delta$ is the Dirac phase.
 The ranges are chosen as $0\leq\theta_{ij}\leq\pi/2$
and $0\leq\delta<2\pi$.
 For Majorana neutrinos,
two additional phases appear, and then the mixing matrix $V$ becomes
\begin{eqnarray}
 V = V_\MNS \times
     \text{diag}( 1, e^{i\varphi_1 /2}, e^{i\varphi_2 /2}),
\end{eqnarray}
where $\varphi_1$ and $\varphi_2$ are referred to as
the Majorana phases~\cite{Schechter:1980gr,Bilenky:1980cx} and
$0 \le \varphi_1,\varphi_2 < 2\pi$.
One has the freedom to work in the basis in which the charged lepton
mass matrix is diagonal, and then the neutrino mass matrix is
diagonalized by $V$.
Using eq.~(\ref{nu_mass}) one can write the couplings
$h_{ij}$ as follows~\cite{Ma:2000wp, Chun:2003ej}:
\begin{equation}
h_{ij}
= \frac{m_{ij}}{\sqrt{2}v_\Delta}
\equiv \frac{1}{\sqrt{2}v_\Delta}
\left[
 V_\MNS
 \text{diag}(m_1,m_2 e^{i\varphi_1},m_3 e^{i\varphi_2})
 V_\MNS^T
\right]_{ij}
\label{hij}
\end{equation}

Neutrino oscillation experiments involving solar~\cite{solar}, 
atmospheric~\cite{atm}, accelerator~\cite{acc},
and reactor neutrinos~\cite{Apollonio:2002gd,:2008ee}
are sensitive to the mass-squared 
differences and the mixing angles, 
and give the following preferred values and ranges:
\begin{eqnarray}
\Delta m^2_{21} \equiv m^2_2 -m^2_1
\simeq 7.6\times 10^{-5} {\rm eV}^2 \,,~~
|\Delta m^2_{31}|\equiv |m^2_3 -m^2_1|
\simeq 2.4\times 10^{-3} {\rm eV}^2\,, \\
\sin^22\theta_{12}\simeq 0.87 \,,~~~~
\sin^22\theta_{23} \simeq 1 \,,~~~~
\sin^22\theta_{13}\lsim 0.14\,.~~~~~~~~~~~~
\label{obs_para}
\end{eqnarray}
We use these values in our numerical analysis
unless otherwise mentioned.
The small mixing angle $\theta_{13}$ has not been measured yet
and hence the value of $\delta$ in completely unknown.
Sensitivity to $\sin^22\theta_{13}\sim 0.01$ is expected 
from various forthcoming
experiments~\cite{futureLBL, futurereac}. Probing
$\sin^22\theta_{13}\ll 0.01$ would require
construction of a neutrino factory
or beta beam experiment~\cite{Bandyopadhyay:2007kx}.
 Since the sign of $\Delta m_{31}^2$ is also undetermined at present, 
distinct neutrino mass hierarchy patterns are possible.
The case with $\Delta m^2_{31} >0$ is referred to as
{\it Normal hierarchy} (NH) where $m_1 < m_2 < m_3$
and the case with $\Delta m^2_{31} <0$ is known as
{\it Inverted hierarchy} (IH) where $m_3 <  m_1 < m_2$.  
 Information on the mass $m_0$ of the lightest neutrino
and the Majorana phases
cannot be obtained from neutrino oscillation experiments.
This is because the oscillation probabilities are independent 
of these parameters, not only in vacuum but also in matter.
If $m_0 \gtrsim 0.2\eV$,
a future ${}^3$H beta decay experiment~\cite{KATRIN}
can measure $m_0$.
Experiments which seek neutrinoless double beta decay%
~\cite{Avignone:2007fu}
are only sensitive to a combination of neutrino masses and phases.
Certainly, extracting information on Majorana phases alone
from these experiments seems extremely difficult, if not impossible%
~\cite{phase-0nbb}.
Therefore it is worthwhile to consider other possibilities in the 
context of the HTM\@.
One method is the branching ratio of the doubly charged Higgs 
boson to two leptons, BR($H^{\pm\pm}\to l^\pm_i l^\pm_j$),
which is determined by $|h_{ij}|^2$ and has been studied in detail in 
\cite{Garayoa:2007fw,Akeroyd:2007zv,Kadastik:2007yd,Perez:2008ha}.
An alternative method is BR($\tlll)$ which depends on 
$|h^*_{\tau i}h_{jk}|^2$ and will be studied in detail in this work.

\section{LFV decays in the HTM}

In this section we introduce the theoretical framework for the
LFV decays $\tlll$, $\meee$ and $\mu\to e \gamma$ in the
HTM\@. Early studies of the effect of $H^{\pm\pm}$ on these decays 
were performed in \cite{Lim:1981kv}.
The works \cite{Chun:2003ej, Kakizaki:2003jk} are the only ones
which address the prediction for such LFV decays in terms of the
parameters of the neutrino mass matrix, which is a unique
feature of the HTM\@. We note here that analogous 
studies of these LFV decays have been performed for the supersymmetric
type II seesaw in which the Higgs triplet has a mass of the 
grand unification (GUT) scale~\cite{Rossi:2002zb, Akhmedov:2008tb}.
In such models
the LFV decays are mediated via TeV scale SUSY particles but the
dependence of $\mu\to e\gamma$ on the neutrino mass matrix parameters is
identical to that for the TeV scale triplet model (HTM) in
\cite{Chun:2003ej, Kakizaki:2003jk}. However, for the LFV 
$\tau$ decays there is a significant difference between models
with a GUT scale triplet and a TeV scale triplet.  
In the former the tree-level contribution to 
$\tlll$
from $H^{\pm\pm}$ is negligible and the dominant contribution 
to these decays is from higher order diagrams. 
Consequently, one expects BR($\tau\to l\gamma)> {\rm BR}(\tlll)$, and the
predictions for $\tau\to e\gamma$ and $\tau\to \mu\gamma$ constitute a probe of 
such models. The distinctive feature of the HTM is the prediction for the
decays $\tlll$ which are mediated by a TeV scale $H^{\pm\pm}$, and
the opposite hierarchy BR($\tlll)> {\rm BR}(\tau\to l\gamma)$.
The LFV processes we study are not mediated by
the neutral scalar triplet fields because $\Delta^0$ does not
couple to charged leptons. In other models (e.g.\ the Two Higgs Doublet Model)
neutral Higgs bosons can contribute
significantly to LFV decays~\cite{Sher:1991km}.

In our numerical analysis the stringent constraint from 
$\meee$ is imposed, while $\tlll$ and $\meg$ offer the possibility
to observe a LFV signal in the HTM\@. Other constraints on $h_{ij}$ 
(e.g.\ the anomalous magnetic moment ($g-2$) of $\mu$, Bhabha scattering and
other LFV processes - reviewed in \cite{Cuypers:1996ia})
are considerably weaker and are neglected. We note that 
$\mu\to e$ conversion in the HTM was also studied in 
\cite{Ma:2000wp, Kakizaki:2003jk} and can give a constraint
similar to that of $\meg$. However, we choose not to impose the constraint from $\mu\to e$ conversion, which
will be considerably weaker than that from $\meee$. In addition, an improvement of the
current bounds \cite{Dohmen:1993mp} on $\mu\to e$ conversion is not likely in the near future
(in contrast to the situation for $\meg$ and $\tlll$).

\subsection{The decays $\tlll$ and $\meee$}

Mere observation of the LFV decays $\tlll$ and $\meee$
would constitute a spectacular signal of physics beyond the SM\@.
In the HTM these decays are mediated at tree level
by $H^{\pm\pm}$ and provide sensitive probes of the $h_{ij}$ couplings 
if $m_{H^{\pm\pm}}$ is of the order of the electroweak scale.
There are six distinct decays for $\tau^- \to \bar{l}_i l_j l_k$
(likewise for $\tau^+$):
$\tau^-\to \mu^+\mu^-\mu^-$, $\tau^-\to e^+e^-e^-$,
$\tau^-\to \mu^+\mu^-e^-$, $\tau^-\to e^+e^-\mu^-$,
$\tau^-\to \mu^+e^-e^-$, $\tau^-\to e^+\mu^-\mu^-$.
Searches for all six decays have been performed by  
BELLE~\cite{Abe:2007ev} and BABAR~\cite{Aubert:2007pw}.
Upper limits of the order BR($\tlll) < 2 \to 8\times 10^{-8}$ were 
obtained. For the decay $\meee$ there exists the stringent bound
BR$(\meee)< 10^{-12}$~\cite{Bellgardt:1987du}.
For a given $m_{H^{\pm\pm}}$ these LFV $\tau$ and $\mu$ decays constrain many combinations of 
the $h_{ij}$ couplings in the HTM\@. Regarding the experimental prospects, 
the sensitivity to BR$(\meee)$ 
will not improve in the foreseeable future and presumably can only
be improved at a $\mu$ storage ring
at a future neutrino factory \cite{Bandyopadhyay:2007kx}.
In contrast, greater sensitivity to $\tlll$ is expected at the 
ongoing $B$ factories, and a proposed Super $B$ factory could probe 
BR($\tlll)< 10^{-9}$ for luminosity $>10$~ab$^{-1}$%
~\cite{Hashimoto:2004sm,Bona:2007qt, Browder:2007gg,Browder:2008em}.
At the LHC, $\tau$ can be copiously produced from several sources
i.e., from $B/D$ decay and direct production via $pp\to W\to \tau\bar{\nu}$,
$pp\to Z\to \tau^+\tau^-$. Sensitivity to BR$(\tmmm)>10^{-8}$ 
is claimed~\cite{Santinelli:2002ea, Giffels:2008ar}, which is similar to
the sensitivity already reached at the $B$ factories.
It is important to note that all the above experiments 
can probe smaller BRs for $\tlll$ than for $\tau\to l\gamma$
and this is due to the smaller SM background for the former.
Hence the hierarchy BR($\tlll)> {\rm BR}(\tau\to l\gamma)$ in the HTM
is favourable from the standpoint of experimental sensitivity.

The virtual exchange of $H^{\pm\pm}$ induces an effective 
interaction of four charged leptons
for $l_m \to \bar{l_i} l_j l_k$ decay ($j\neq k$) as follows:
\beqa
{\cal L}
&=&
\frac{ (h^\ast)_{mi}(h)_{jk} }{ 4\sqrt{2}\,G_F m_{H^{\pm\pm}}^2}
\left\{
 2\sqrt{2}\,G_F
 \left( \overline l_m\gamma^\mu P_Ll_{k} \right)
 \left( \overline l_i\gamma_\mu P_Ll_{j} \right)
\right\}\nonumber\\
&&\hspace*{20mm}
{}+
 (\text{three terms with $m\leftrightarrow i$ and $j\leftrightarrow k$})
+ h.c.
\ .
\eeqa
 Here $m, i, j, k$ are fixed and we used the Fierz transformation.
 $G_F = 1.17\times 10^{-5}\,\GeV^{-2}$ is the Fermi constant.
Neglecting the masses of the final-state particles,
the branching ratio for $\tlll$ is given by:
\begin{eqnarray}
\BR(\tlll)
&=&
 \frac{\ S\,|h_{\tau i}|^2 |h_{jk}|^2 }
      { 4 G_F^2 m_{H^{\pm\pm}}^4 }\,
 \BR(\tau\to \mu\overline{\nu}\nu)\\
&\simeq&
 0.19\,S\,|h_{\tau i}|^2 |h_{jk}|^2
 \left( \frac{ 200\GeV }{ m_{H^{\pm\pm}} } \right)^4 ,
\label{BRtaulll}
\end{eqnarray}
where $\BR(\tau\to \mu\overline{\nu}\nu) \simeq 17\%$.
Here $S$=1 (2)  for $j=k$ ($j\ne k$). 
The branching ratio for $\meee$ is given by:
\begin{eqnarray}
\BR(\meee)
&=&
 \frac{\ |h_{\mu e}|^2 |h_{ee}|^2 }{ 4 G_F^2 m_{H^{\pm\pm}}^4 }\,
 \BR(\mu\to e\overline{\nu}\nu)\\
&\simeq&
 1.1\,|h_{\mu e}|^2 |h_{ee}|^2
 \left( \frac{ 200\GeV }{ m_{H^{\pm\pm}} } \right)^4\ .
\label{BRmueee}
\end{eqnarray}
where BR$(\mu\to e\overline{\nu}\nu)\simeq 100\%$.

Previous studies of the above decays 
in the context of the HTM were performed
in \cite{Chun:2003ej, Kakizaki:2003jk}.
In \cite{Chun:2003ej}
only the specific case of $|h_{e\mu}|=0$ was studied, which
automatically suppresses BR($\meee$), and predictions for the ratios
of BR$(\tlll)$ were given.
In \cite{Kakizaki:2003jk}
it was mentioned that an observable BR($\tlll$) would require
accidental cancellations to suppress BR($\meee$),
although there was no quantitative analysis.

We note that studies of $\tlll$ (and $\meee$)
in other models which contain $H^{\pm\pm}$
have been performed  (hereafter the subscripts L and R indicate the chirality of
the leptons in the interaction of eq.~(\ref{trip_yuk}), and $H^{\pm\pm}$ in the 
HTM corresponds to $H^{\pm\pm}_L$): 
i) Zee-Babu model ($H^{\pm\pm}_R$ only) 
\cite{Babu:2002uu}; 
ii)
Left-Right symmetric model ($H^{\pm\pm}_L$ and $H^{\pm\pm}_R$)%
~\cite{Cirigliano:2004mv, Akeroyd:2006bb}; iii) 
other models which contain a $H^{\pm\pm}$~\cite{Chang:2005ag}.
 Analyses using effective Lagrangians have been performed%
~\cite{Dassinger:2007ru} and can be applied to models with $H^{\pm\pm}$.
Angular asymmetries
can also be defined which can probe the relative strength 
of the contributions from $H^{\pm\pm}_L$ and $H^{\pm\pm}_R$%
~\cite{Giffels:2008ar, Akeroyd:2006bb, Kitano:2000fg}.
In the HTM such an asymmetry would be maximal since $H^{\pm\pm}$ only interacts
with left-handed leptons.
The distinctive phenomenological feature of $H^{\pm\pm}$ in the HTM is the
specific relationship between $h_{ij}$ and the neutrino mass matrix
given by eq.~(\ref{nu_mass}), which is not realized in the above models.

\subsection{$\mu \to e\gamma$}
Searches for $\mu \to e\gamma$ have a long history 
(e.g., see \cite{Kuno:1999jp}), 
and the most stringent upper bound
is from the MEGA Collaboration which obtained BR($\mu \to e\gamma)< 1.2\times 10^{-11}$
\cite{Brooks:1999pu}. The ongoing MEG experiment anticipates sensitivity to 
BR($\mu \to e\gamma) \sim 10^{-13}$ \cite{Grassi:2005ac}.
The effective Lagrangian for $\mu \to e \gamma$ is as follows:  
\begin{eqnarray}
{\cal L}=-2\sqrt{2}\,G_F \left\{m_\mu A_R\overline\mu\sigma^{\mu\nu}P_Le F_{\mu\nu}
+m_\mu A_L\overline\mu \sigma^{\mu\nu}P_Re F_{\mu\nu}+h.c. 
\right\} \ ,
\end{eqnarray}
where $\sigma^{\mu\nu}=i[\gamma^\mu, \gamma^\nu]/2$.
In the HTM $A_L=0$ but $A_R$ receives contributions from $H^{\pm\pm}$ and $H^\pm$.
Explicit expressions for $A_R$ have been obtained, e.g.,
following \cite{Lavoura:2003xp}:
\begin{eqnarray}
A_R
\simeq -\frac{ q_e (h^\dagger h)_{e\mu} }{ 48\sqrt{2}\,\pi^2 G_F }
     \left(
      \frac{1}{m_{H^{\pm\pm}}^2}
      + \frac{1}{8 m_{H^\pm}^2}
     \right)\ ,
\label{eq:ar}
\end{eqnarray}
where $q_e$ is the positron charge and we have neglected 
the electron mass in the final state and all lepton masses in the loop.
If $m_{H^{\pm\pm}}\simeq m_{H^{\pm}}$ 
the dominant contribution is from the loops involving virtual $H^{\pm\pm}$. The decay rate 
is determined by the combination $(h^\dagger h)_{e\mu}$, 
and the branching ratio for $\mu \to e\gamma$
with $m_{H^\pm} \simeq m_{H^{\pm\pm}}$ is given by:
\footnote{In our numerical analysis we do not  
include a suppression factor of
$\sim 15\%$ arising from electromagnetic
corrections \cite{Czarnecki:2001vf}.}
\begin{eqnarray}
\BR(\mu\to e\gamma) \simeq 384\pi^2 |A_R|^2
&\simeq&
 \frac{ 27 \alpha |(h^\dagger h)_{e\mu}|^2 }
      { 64 \pi G_F^2 m_{H^{\pm\pm}}^4 }\\
&\simeq&
 4.5\times 10^{-3}\,|(h^\dagger h)_{e\mu}|^2
 \left( \frac{ 200\GeV }{ m_{H^{\pm\pm}} } \right)^4\ .
\end{eqnarray}
Here $\alpha \equiv q_e^2/4\pi = 1/137$ is the fine structure constant.
It was noted in \cite{Chun:2003ej,Kakizaki:2003jk}
that the decay rate for  $\mu\to e\gamma$ is very sensitive to
$s_{13}$, but it is not so sensitive
to the neutrino mass spectrum
because it is independent of the absolute neutrino mass.
Importantly (and stressed in \cite{Akhmedov:2008tb}),
the coupling $(h^\dagger h)_{e\mu}$
has no dependence on the Majorana phases either.
Hence  the prediction for BR$(\mu \to e\gamma)$
in the HTM is much sharper than that for $\tlll$,
but the former is unable to probe those
neutrino parameters which cannot be probed in neutrino 
oscillation experiments (i.e., absolute neutrino
mass and Majorana phases).
If polarized muons are available, then an angular asymmetry 
can be defined for $\mu \to e\gamma$ \cite{Okada:1999zk}.
In the HTM such an asymmetry would be maximal \cite{Akeroyd:2006bb}
because $H^{\pm\pm}$
and $H^\pm$ only interact with left-handed leptons.


\section{Numerical Results}
In this section we present our numerical results for the 
BRs of the decays $\tlll$ and $\mu \to e\gamma$ as a function of the 
parameters of the neutrino mass matrix.
In eq.~(\ref{hij}) one can express $m_1,m_2,m_3$ in terms of 
two neutrino mass-squared differences ($\Delta m^2_{21},\Delta m^2_{31}$)
and the mass of the lightest neutrino $m_0$.
The couplings $h_{ij}$ are functions of nine parameters:\\
$\Delta m^2_{21}$,$\Delta m^2_{31}$,$m_0$, 
three mixing angles $(\theta_{12},\theta_{13},\theta_{23})$, 
and three complex phases $(\delta, \varphi_1, \varphi_2)$.

The work of \cite{Chun:2003ej} focussed on the case of suppressing
BR($\meee$) by small $|h_{e\mu}|$. For the special case of
$|h_{e\mu}|=0$ the ratios of $\tlll$ and 
$\mu\to e\gamma$ were studied, normalizing
to a specific decay $\tlll$. 
In analogy with the analysis of \cite{Chun:2003ej} our numerical analysis will study
the values of ratios of LFV decays in which the arbitrary triplet vacuum expectation
value cancels out. However, in difference to \cite{Chun:2003ej}, some of our
numerical analysis will normalize to
BR$(\meee)$ (which is not set to zero) and will quantify the parameter space 
of the neutrino mass matrix where a signal for $\tlll$ and/or 
$\mu\to e\gamma$ could be seen at ongoing and planned experiments. 
The condition
\begin{eqnarray}
\frac{\BR(\tlll)}{\BR(\meee)} > 10^{3}
\label{cond:lll}
\end{eqnarray}
signifies that BR$(\tlll) > 10^{-9}$ is possible
while satisfying the bound BR$(\meee) < 10^{-12}$.
 BR$(\tlll) > 10^{-9}$ is within the 
sensitivity of proposed high luminosity $B$ factories.
Furthermore,
$\ratio{\tlll}{\meee}> 10^{4}$
signifies that $\tlll$
can be observed even in the current runs of the $B$ factories
with integrated luminosities of the order of 1 ${\rm ab}^{-1}$.
The condition
\begin{eqnarray}
\frac{\BR(\mu \to e\gamma)}{\BR(\meee)} > 10^{-1}
\label{cond:meg}
\end{eqnarray}
corresponds to BR$(\mu \to e\gamma)>10^{-13}$, 
which is within the sensitivity of the MEG experiment.
 These are {\sl necessary} but not
sufficient conditions for observation of
$\tlll$ and/or $\meg$ in the HTM\@.

Once $\BR(\meee)$ is sufficiently suppressed
to satisfy both eq.~(\ref{cond:lll}) and (\ref{cond:meg}),
there is the possibility of observing multiple
LFV signals while respecting current bounds on 
$\tlll$ and $\mu \to e\gamma$ if the following conditions are satisfied:
\begin{eqnarray}
10^2 <
 &{\displaystyle \frac{\BR(\tlll)}{\BR(\mu \to e\gamma)}}& ,
\label{cond:meg-2}\\
&{\displaystyle \frac{\BR(\tau\to\bar{l}_a l_b l_c)}{\BR(\mu \to e\gamma)}}&
 < 10^5 ,
\label{cond:meg-3}
\end{eqnarray}
where $\BR(\tau\to\bar{l}_a l_b l_c)$ is the largest
among $\BR(\tlll)$.
 For the ratio $< 10^2$ in eq.~(\ref{cond:meg-2})
only a signal of $\mu \to e\gamma$ is possible,
corresponding to $\BR(\tlll) \lesssim 10^{-9}$
with the current bound $\BR(\mu\to e\gamma) \lesssim 10^{-11}$
even if eq.~(\ref{cond:lll}) is satisfied.
 Only $\tlll$ can be observed for the ratio
$> 10^5$ in eq.~(\ref{cond:meg-3}),
for which the current bound $\BR(\tau\to\bar{l}_a l_b l_c) \lesssim 10^{-8}$
gives $\BR(\mu\to e\gamma) \lesssim 10^{-13}$
even if eq.~(\ref{cond:meg}) is satisfied.
 If more than two $\BR(\tlll)$ satisfy eq.~(\ref{cond:meg-2}),
a condition for the ratio of $\tau$ LFV decays
\begin{eqnarray}
 \frac{\BR(\tlll)}{\BR(\tau\to\bar{l}_a l_b l_c)} > 10^{-1}
\label{cond:lll-2}
\end{eqnarray}
makes it possible to observe $\tlll$ in
addition to $\tau\to\bar{l}_a l_b l_c$.
The above ratios will be used to interpret our numerical results and
find the regions of parameter space for possible LFV signals.
Note again that conditions (\ref{cond:meg-2})-(\ref{cond:lll-2})
are {\sl necessary} but not sufficient.
Once these conditions are satisfied, the arbitrary 
$v_\Delta$ and $m_{H^{\pm\pm}}$ can be freely chosen to provide an observable 
BR because such conditions do not depend on these parameters.

We present below explicit expressions for $h_{ee}$ and $h_{e\mu}$ (which determine the 
decay rate for $\mu\to eee$), and
the remaining $h_{ij}$ can be found in 
\cite{Garayoa:2007fw, Akeroyd:2007zv, Kadastik:2007yd, Perez:2008ha}.
\begin{eqnarray}
h_{ee} &=& \frac{1}{\sqrt{2} v_\Delta}
 \Bigl(
  m_1 c_{12}^2 c_{13}^2
  + m_2 s_{12}^2 c_{13}^2 e^{i\varphi_1}
  + m_3 s_{13}^2 e^{i(\varphi_2-2\delta)}
 \Bigr)\,, \\
  \label{hee}
h_{e\mu} &=& \frac{1}{\sqrt{2} v_\Delta}
 \Bigl\{
  m_1 ( -s_{12}c_{23} - c_{12}s_{23}s_{13} e^{i\delta} ) c_{12}c_{13}
 \nonumber \\
 &&\hspace{27mm}
 {}+ m_2 ( c_{12}c_{23} - s_{12}s_{23}s_{13} e^{i\delta} ) s_{12}c_{13} e^{i\varphi_1}
  + m_3 s_{23}c_{13}s_{13} e^{-i\delta} e^{i\varphi_2}
 \Bigr\}\,,
\label{hij_expressions}
\end{eqnarray}
Four cases corresponding to no CP violation from Majorana phases 
can be defined as follows: 
Case~I $(\varphi_1=0,\varphi_2=0)$;
Case~II $(\varphi_1=0,\varphi_2=\pi)$;
Case~III $(\varphi_1=\pi,\varphi_2=0)$; 
Case~IV $(\varphi_1=\pi,\varphi_2=\pi)$.
These four cases have been studied
in \cite{Chun:2003ej} for values
of $m_0=0$ or {$\cal O$}(1)~eV\@. In this work we will study in detail the 
dependence of $\tlll$ and $\mu\to e\gamma$ on the neutrino 
mass matrix parameters (in particular $s_{13}$, $\delta$, $m_0$, 
$\varphi_1$ and $ \varphi_2$) in order to find the regions
which can provide an observable signal at the $B$ factories
(for $\tlll$) and the MEG experiment (for  $\mu\to e\gamma$).
We consider two distinct cases for the necessary suppression of
$\meee$, corresponding to $|h_{e\mu}|\simeq 0$ and
 $|h_{ee}|\simeq 0$.

\subsection{Case of $|h_{e\mu}|\simeq 0$}
\subsubsection{Normal Hierarchy}

In the scenarios with no CP violation from Majorana phases,
the special case of $|h_{e\mu}|=0$ is achieved at a specific (``magic'')
value of $\theta_{13}$ for all values of $m_1$ ($=m_0$).
 In the normal hierarchy,
the magic value is given by
\begin{eqnarray}
\text{Case I}\ (\varphi_1=\varphi_2=0) &:&
\displaystyle
 s_{13}^\magic
  = \frac{ (m_2-m_1) c_{23} \sin{2\theta_{12}} }
         { 2 ( -m_1 c_{12}^2 - m_2 s_{12}^2 + m_3 ) s_{23} }, \
 \delta = \pi,
\label{eq:s13mgc-1}\\
\text{Case II}\ (\varphi_1=0, \varphi_2=\pi) &:&
\displaystyle
 s_{13}^\magic
  = \frac{ (m_2-m_1) c_{23} \sin{2\theta_{12}} }
         { 2 ( m_1 c_{12}^2 + m_2 s_{12}^2 + m_3 ) s_{23} }, \
 \delta = 0,
\label{eq:s13mgc-2}\\
\text{Case III}\ (\varphi_1=\pi, \varphi_2=0) &:&
\displaystyle
 s_{13}^\magic
 = \frac{ (m_1+m_2) c_{23} \sin{2\theta_{12}} }
        { 2 ( -m_1 c_{12}^2 + m_2 s_{12}^2 + m_3 ) s_{23} }, \
 \delta = 0,
\label{eq:s13mgc-3}\\
\text{Case IV}\ (\varphi_1=\varphi_2=\pi) &:&
\displaystyle
 s_{13}^\magic
 = \frac{ (m_1+m_2) c_{23} \sin{2\theta_{12}} }
        { 2 ( m_1 c_{12}^2 - m_2 s_{12}^2 + m_3 ) s_{23} }, \
 \delta = \pi,
\label{eq:s13mgc-4}
\end{eqnarray}
where we define $s_{13}^\magic \equiv \sin{\theta_{13}^\magic}$
and the value of $\delta$ is chosen to give $|h_{e\mu}|=0$.
 For all four cases,
$\sin^2{2\theta_{13}^\magic}$ is about 0.02 for $m_1=0$; for Case~I and II
$\sin^2{2\theta_{13}^\magic}$ is a decreasing function
of $m_1$ and converges to $9\times 10^{-4}$ and 0, respectively.
 On the other hand,
$\sin^2{2\theta_{13}^\magic}$ for Case~III and IV
is an increasing function of $m_1$,
and then $\sin^2{2\theta_{13}} < 0.14$ can be satisfied
only for $m_1 \lesssim 0.01\eV$.
If $\sin^2{2\theta_{13}}$ is tuned completely to the magic value,
the constraint from $\meee$ is automatically satisfied
and one cannot normalize BRs to BR($\meee$)~\cite{Chun:2003ej}.

\begin{figure}[t]
\begin{center}
\includegraphics[origin=c, angle=-90, scale=0.3]{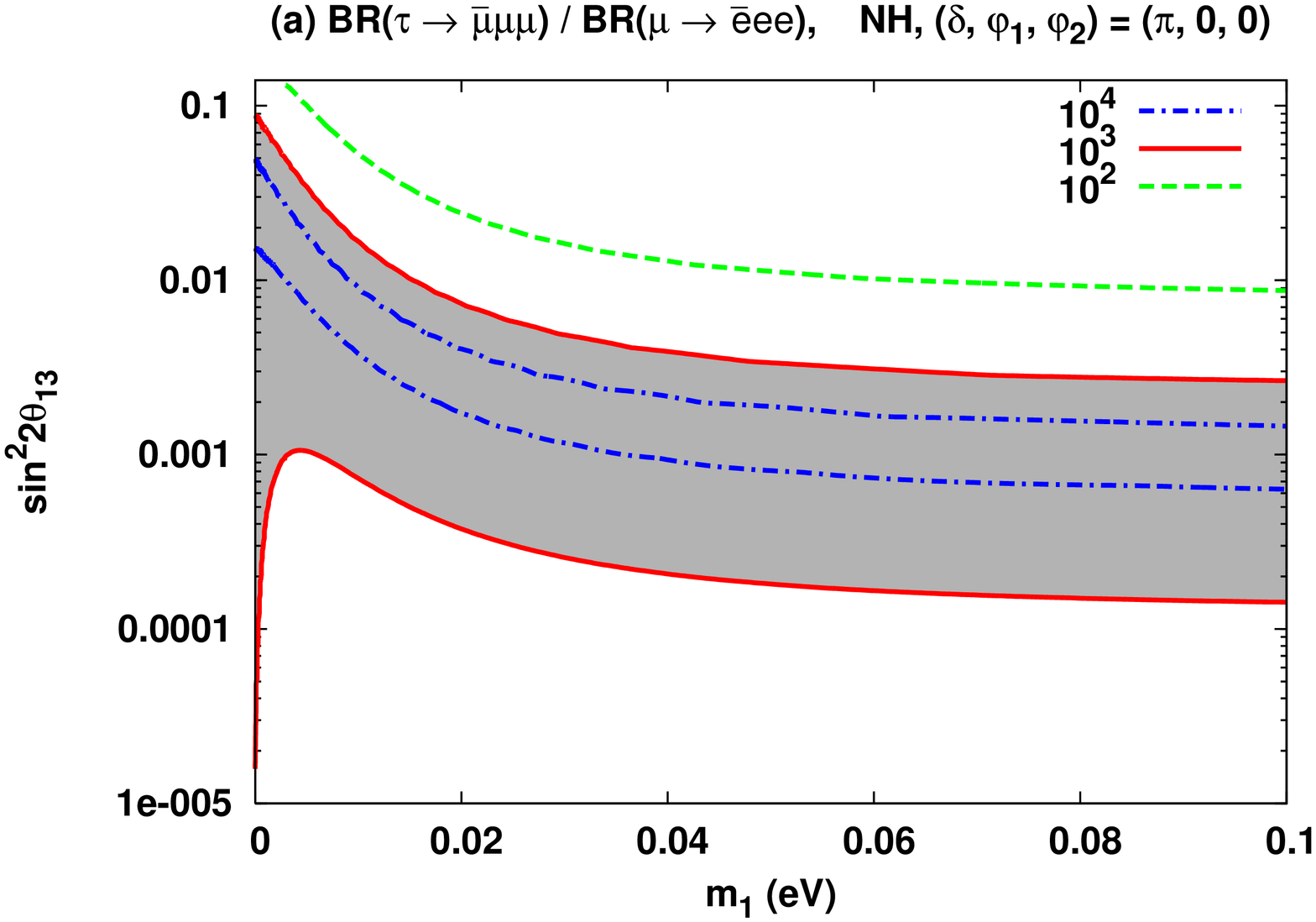}
\includegraphics[origin=c, angle=-90, scale=0.3]{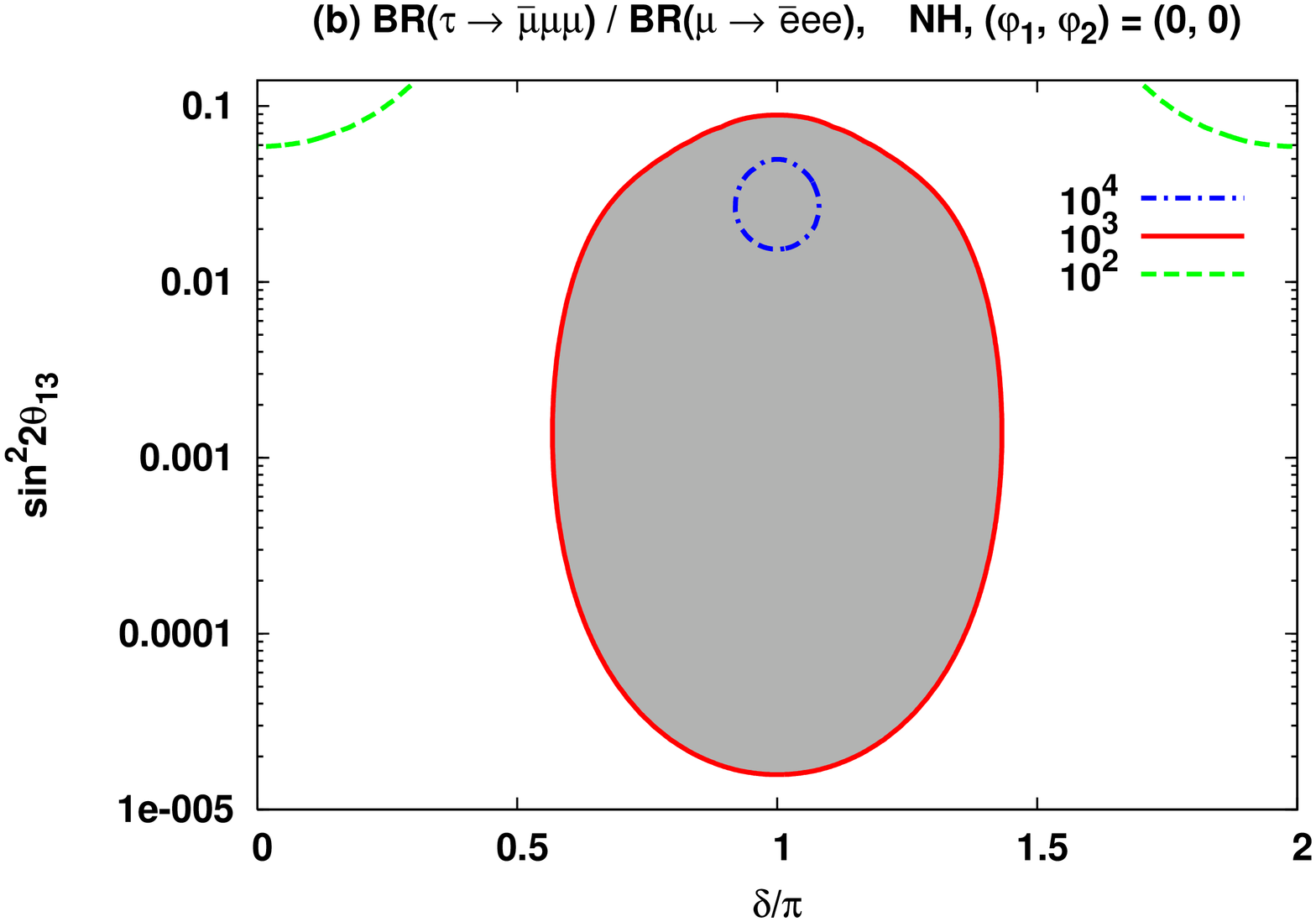}
\vspace*{-10mm}
\caption{
Contours of the ratio \ratio{\tmmm}{\meee}
in Case~I in the normal hierarchy.
 a) $m_1$ dependence for $\delta = \pi$.
For any $m_1$, vanishing $|h_{e\mu}|$ is achieved
at the magic value of $s_{13}$ from eq.~(\ref{eq:s13mgc-1}). 
 b) $\delta$ dependence with $m_1=0$.
 A signal for $\tmmm$ is possible if the ratio $> 10^3$ (depicted by the shaded region).
 }
\label{fig:tmmm_hem0_n}
\end{center}
\end{figure}

In the following
we discuss in detail the values of ratios of LFV decays
in the case of eq.~(\ref{eq:s13mgc-1}).
In this case
$\BR(\tmmm)$ is the largest one among $\BR(\tlll)$.
Fig.~\ref{fig:tmmm_hem0_n} shows contours of the ratio \ratio{\tmmm}{\meee}.
This ratio is displayed in the plane $[m_1, \sin^22\theta_{13}]$
in Fig.~\ref{fig:tmmm_hem0_n}(a).
It is evident that a wide range of $\sin^2{2\theta_{13}}$
around its magic value for small $m_1$
gives a ratio $>10^3$,
which is the necessary condition to provide a signal of $\tmmm$
at high luminosity $B$ factories.
This can be traced to the fact that
BR($\meee$) depends on the second power of the coupling $|h_{e\mu}|$, 
and so obtaining a ratio $>10^3$ does not
constitute a fine-tuning when expressed in terms of the neutrino
parameters which determine $h_{e\mu}$.
In Fig.~\ref{fig:tmmm_hem0_n}(b) the ratio is displayed
in the plane $[\delta, \sin^22\theta_{13}]$ with $m_1=0$.
One sees that achieving a ratio $> 10^3$ 
can be obtained in the interval $0.5 < \delta/\pi < 1.5$ for a 
wide range of $\sin^2{2\theta_{13}}$ around its magic value.
Since only the relative Majorana phase $\varphi_2 - \varphi_1$
is physical for $m_1 = 0$,
Fig.~\ref{fig:tmmm_hem0_n}(b) is also the result for Case~IV\@.
 Roughly speaking,
the region of large ratio in Fig.~\ref{fig:tmmm_hem0_n}(b)
shifts horizontally for different values of the Majorana phases.
Consequently, the
$\delta$-dependence for Case~II and III is similar to
the result in Fig.~\ref{fig:tmmm_hem0_n}(b)
with a shift of $\delta$ by $\pi$.
 If $m_1$ is not very small,
$|h_{e\mu}|\simeq 0$ with $s_{13}^\magic$ of eq.~(\ref{eq:s13mgc-1})
requires considerable fine-tuning of three phases.
 Even for large $m_1$,
$s_{13}^\magic$ of eq.~(\ref{eq:s13mgc-2})
gives $|h_{e\mu}|\simeq 0$ in a sizeable region
of $\delta$ and $\varphi_2$.

\begin{figure}[t]
\begin{center}
\includegraphics[origin=c, angle=-90, scale=0.3]{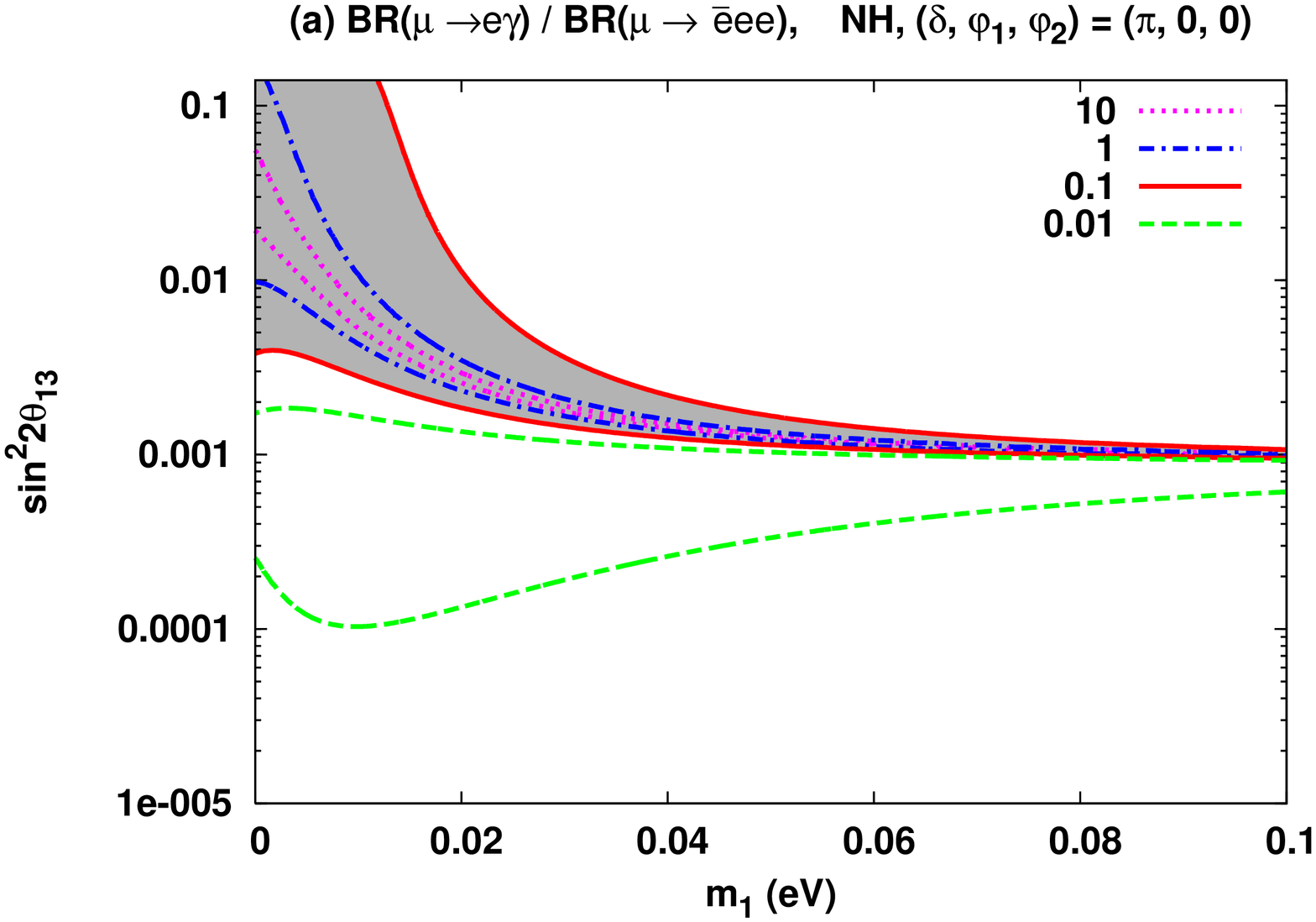}
\includegraphics[origin=c, angle=-90, scale=0.3]{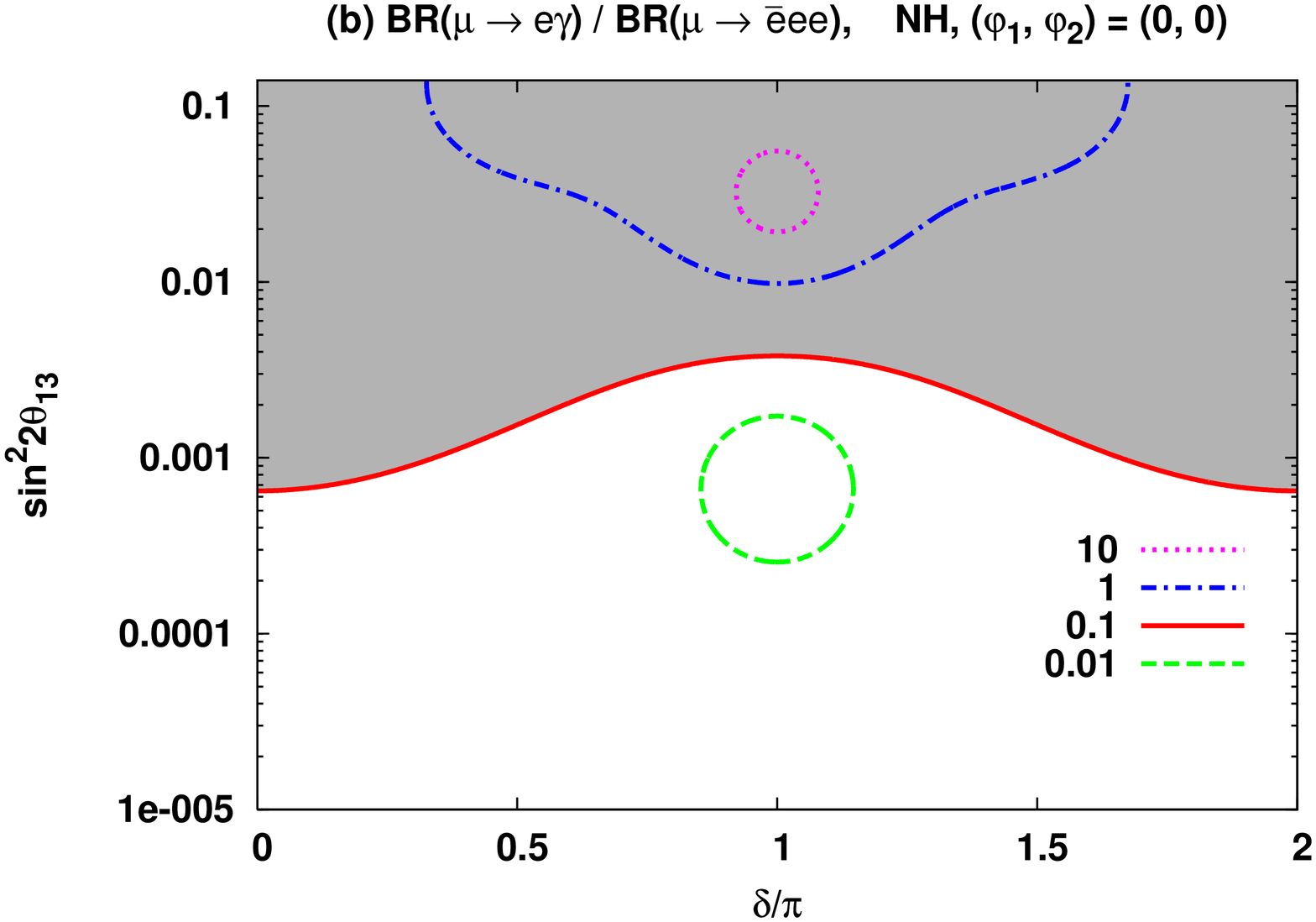}
\vspace*{-10mm}
\caption{
Contours of the ratio \ratio{\mu\to e\gamma}{\meee}
in Case~I in the normal hierarchy.
 a) $m_1$ dependence for $\delta = \pi$.
For any $m_1$, vanishing $|h_{e\mu}|$ is achieved
at the magic value of $s_{13}$ from eq.~(\ref{eq:s13mgc-1}).
 b) $\delta$ dependence for $m_1=0$.
 A signal for $\mu\to e\gamma$ is possible if the ratio $> 0.1$
(depicted by the shaded region).
}
\label{fig:meg_hem0_n}
\end{center}
\end{figure}

\begin{figure}[t]
\begin{center}
\includegraphics[origin=c, angle=-90, scale=0.3]{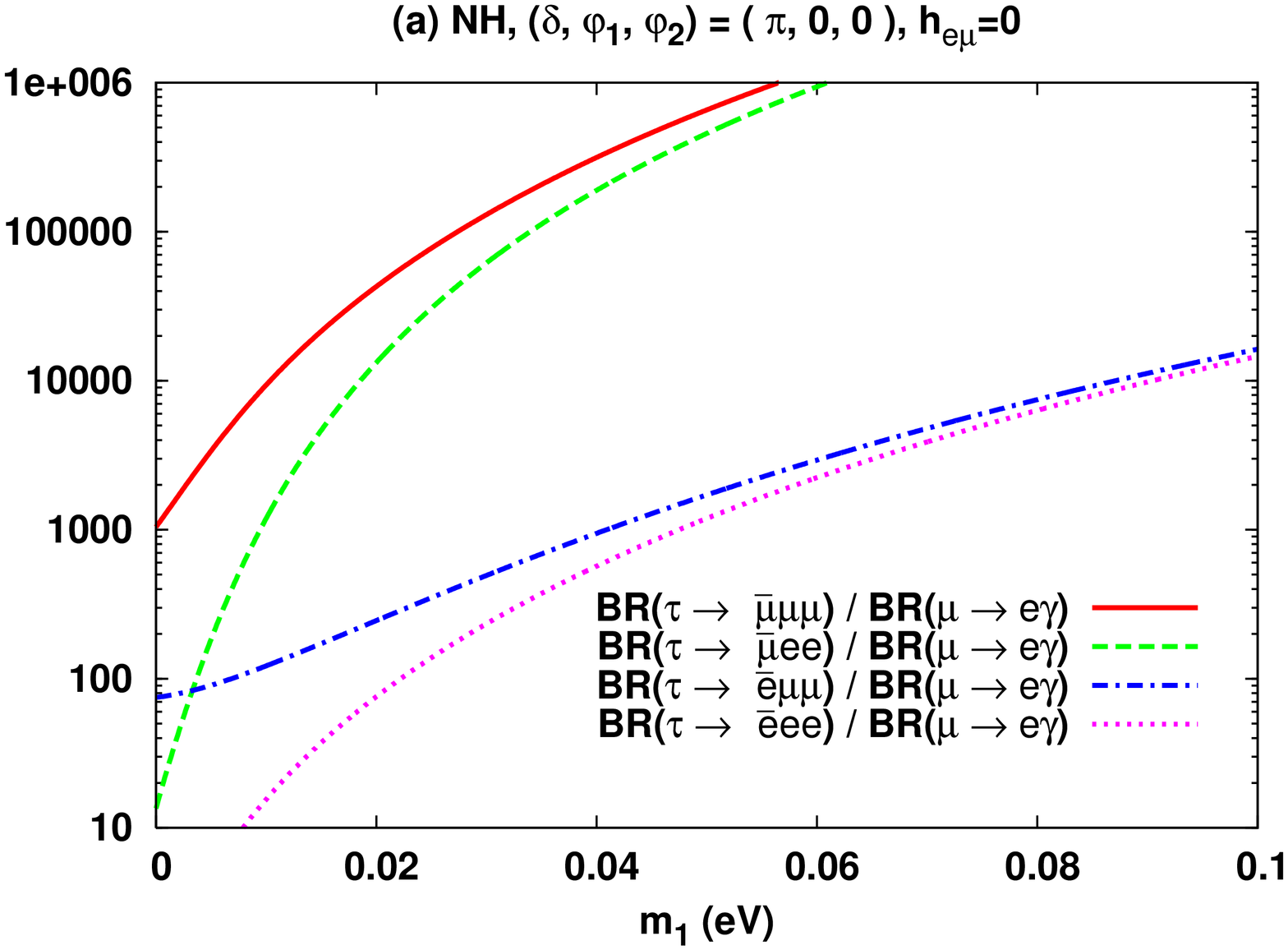}
\includegraphics[origin=c, angle=-90, scale=0.3]{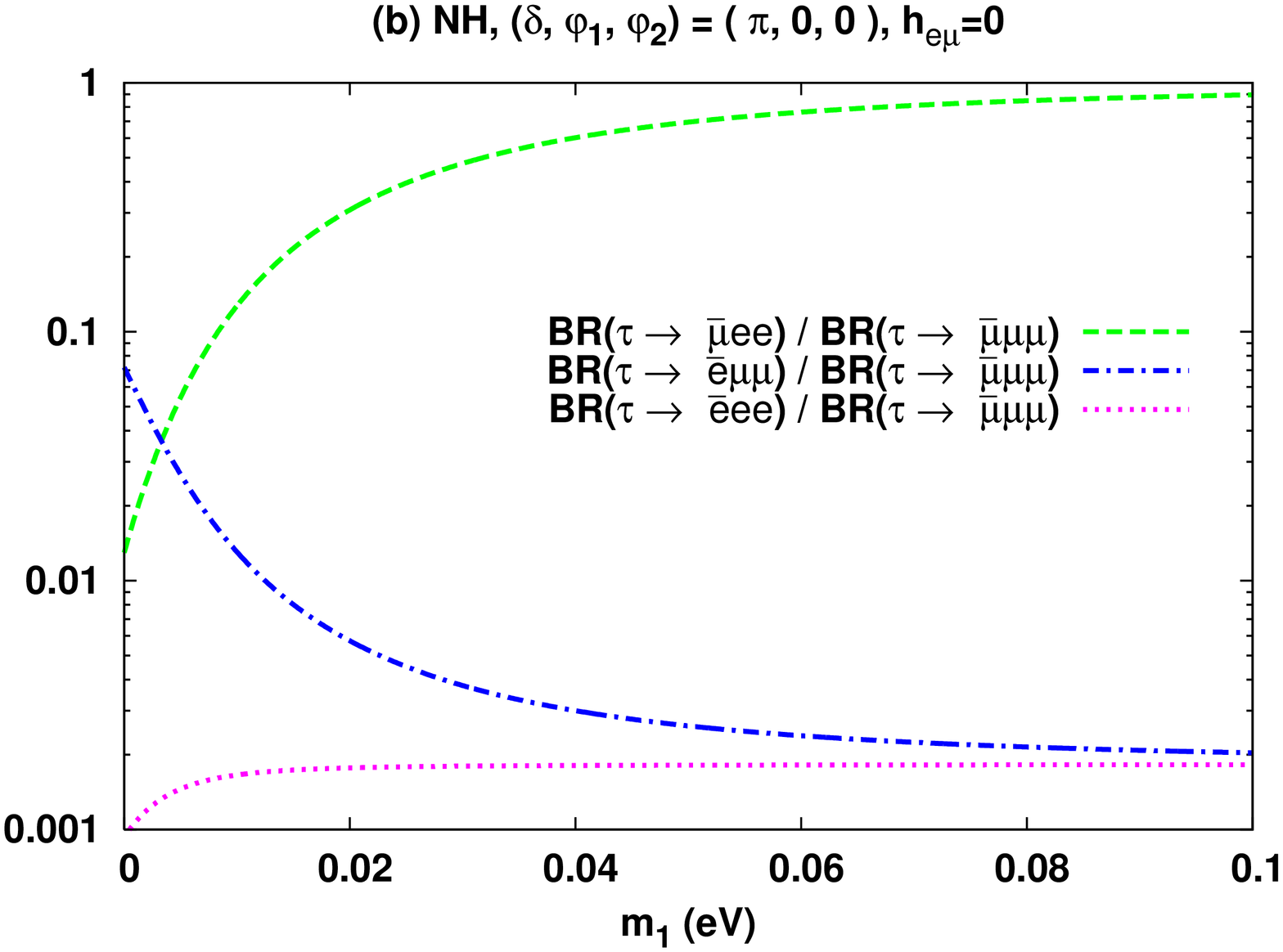}
\vspace*{-10mm}
\caption{
 $m_1$-dependence of ratios of LFV decay branching ratios
for $|h_{e\mu}|=0$ with $s_{13}^\magic$ and $\delta = \pi$
in Case~I in the normal hierarchy. 
 a) The solid, dashed, dash-dotted, and dotted lines show
results for
$\BR(\tmmm)$, $\BR(\tmee)$,
$\BR(\tau\to\bar{e}\mu\mu)$, and $\BR(\tau\to\bar{e}ee)$
respectively.
BRs are normalized by $\BR(\meg)$.
See also the conditions (\ref{cond:meg-2}) and (\ref{cond:meg-3}).
 b) Branching ratios of LFV $\tau$ decays
normalized by $\BR(\tmmm)$.
See also the condition (\ref{cond:lll-2}).
}
\label{fig:ratio_hem0_n}
\end{center}
\end{figure}

 In Fig.~\ref{fig:meg_hem0_n} the ratio
\ratio{\mu \to e\gamma}{\meee} is plotted for Case~I\@.
 Fig.~\ref{fig:meg_hem0_n}(a)
shows that a ratio $>10^{-1}$,
which is necessary for discovery of $\mu\to e\gamma$,
is obtained without fine-tuning of $\sin^2{2\theta_{13}}$
to the magic value for small $m_1$.
In contrast, for large $m_1$
fine-tuning seems to be necessary for a ratio $>10^{-1}$.
Since $\BR(\mu\to e\gamma)$ does not depend on
the absolute neutrino mass,
the behaviour in Fig.~\ref{fig:meg_hem0_n}(a) is a result of
the $m_1$ dependence of $\BR(\meee)$. We will see below 
(Fig.~\ref{fig:ratio_hem0_n})
that $\mu\to e\gamma$ cannot be observed in the region of large $m_1$
even if $\theta_{13}$ is fine-tuned to $\theta_{13}^\magic$ since
current limits on LFV $\tau$ decays would be violated.
The ratio becomes very small at $\sin^2{2\theta_{13}}\simeq 0.001$
because $|(hh^\dagger)_{e\mu}|\simeq 0$. Fig.~\ref{fig:meg_hem0_n}(b)
shows the $\delta$-dependence of the ratio for $m_1=0$.
Most of the plane above $\sin^2{2\theta_{13}}\simeq 10^{-3}$
allows the ratio $>10^{-1}$, and
the large ratio is achieved at any value of $\delta$.
The circle of ratio $<10^{-2}$ in Fig.~\ref{fig:meg_hem0_n}(b)
exists around the point of $|(hh^\dagger)_{e\mu}| = 0$,
which does not depend on Majorana phases. 
Therefore the ratio is small for
$\sin^2{2\theta_{13}} \lesssim 10^{-3}$
 even with different values of Majorana phases.
Comparing Fig.~\ref{fig:tmmm_hem0_n} with Fig.~\ref{fig:meg_hem0_n}
one can see the parameter regions
which allow a signal for one or both of $\tmmm$ and $\mu\to e\gamma$. 
Most notably, a signal for at least one LFV decay is possible
in a large part of the parameter space for small $m_1$.
In the region where signals for both $\tmmm$ and $\mu\to e\gamma$
are possible, the current bounds on these decays must also be respected.

 Once $s_{13}^\magic$ is taken,
$\ratio{\tlll}{\meee} > 10^3$
and $\ratio{\mu \to e\gamma}{\meee} >10^{-1}$
are satisfied automatically.
 Then current bounds on BR($\tlll$) and BR($\mu\to e\gamma$) should be considered.
 Fig.~\ref{fig:ratio_hem0_n} shows the $m_1$-dependence
of several ratios of LFV BRs normalized to $\BR(\mu\to e\gamma)$
with $\delta=\pi$ and $s_{13}=s_{13}^\magic$
of eq.~(\ref{eq:s13mgc-1}) for Case~I\@. 
Both $\tau\to \bar{\mu}\mu e$
and $\tau\to \bar{e}\mu e$ vanish for $s_{13}=s_{13}^\magic$.
For $m_1 \gtrsim 0.03\,\eV$, the ratio for  $\tmmm$ exceeds $10^5$ 
and so $\mu\to e\gamma$ is difficult to observe.
Although Fig.~\ref{fig:ratio_hem0_n}(a)
shows that $\ratio{\tlll}{\mu\to e\gamma} > 10^{2}$
is satisfied for several $\tlll$ (e.g.\
all four $\tlll$ decays satisfy this condition for $m_1 \sim 0.03\,\eV$)
one can see in Fig.~\ref{fig:ratio_hem0_n}(b)
that only the decay $\tau\to\bar{\mu}ee$ can satisfy
$\ratio{\tlll}{\tmmm} > 10^{-1}$
and be observed in addition to $\tmmm$ for $m_1 \gtrsim 0.01\,\eV$.
For $m_1 \lesssim 0.01\,\eV$, $\tmmm$ is the only LFV $\tau$ decay which can be observed.
 Table~\ref{tab:LFV_hem0} summarizes the LFV decays
which can be measured for the case with eq.~(\ref{eq:s13mgc-1}).

\begin{table}[t]
\begin{tabular}{c||c|c|c}
 & \multicolumn{3}{|c}{$(\delta, \varphi_1, \varphi_2) = (\pi, 0, 0)$}\\
\cline{2-4}
 &\, $m_1 \lesssim 0.01\,\eV$\,
 &\, $0.01\,\eV \lesssim m_1 \lesssim 0.03\,\eV$\,
 &\, $0.03\,\eV\lesssim m_1$\\
\hline\hline
NH \
 & $\tmmm$
 & $\tmmm$
 & $\tmmm$\\[-3mm]
 &
 & $\tmee$
 & $\tmee$\\[-3mm]
 & $\meg$
 & $\meg$
 &\\
\hline
IH \
 & \multicolumn{3}{|c}{$\tmmm$}
\\[-3mm]
 & \multicolumn{3}{|c}{$\tmee$}
\end{tabular}
\caption{
Table of LFV decays which can be
within the future experimental sensitivity with $|h_{e\mu}|=0$
for $\delta = \pi$ in Case~I\@.
Other decay modes are rather unlikely to be observed.
}
\label{tab:LFV_hem0}
\end{table}

\subsubsection{Inverted Hierarchy}
 In the inverted hierarchy scenario
the condition for $|h_{e\mu}|= 0$ differs from that in the 
hierarchical scenario, as was noted in \cite{Chun:2003ej}.
In Case~I and II,  $|h_{e\mu}|=0$ is achieved when $\delta =0$ 
with a magic value of $s_{13}$,
\begin{eqnarray}
 s_{13}^\magic
 = \frac{ (m_2-m_1) c_{23} \sin{2\theta_{12}} }
        { 2 ( m_1 c_{12}^2 + m_2 s_{12}^2 \mp m_3 ) s_{23} },
\label{eq:s13mgcIH}
\end{eqnarray} 
where the upper and lower signs of the $m_3 (=m_0)$ term
are for Case~I and II respectively.
At $m_3=0$ the value of
$\sin^2{2\theta_{13}^\magic}$ is $2\times 10^{-4}$ 
and converges to $9\times 10^{-4}$ (0) for large $m_3$
in Case~I (II).
 No acceptable $s_{13}^\magic$ exists for Case~III and IV\@.
 In the inverted hierarchy,
BR($\tmee$) is the largest one among BR($\tlll$).
 Only a small region of $s_{13}$ can give
$\ratio{\tmee}{\meee} > 10^{3}$ for Case~I in IH\@.
Fine tunings of three phases are required except for very small $m_3$.
The parameter space in the plane $[m_1,\sin^22\theta_{13}]$
for $\ratio{\meg}{\meee} > 0.1$
is similar to that for $\ratio{\tmee}{\meee} > 10^3$.
Taking $s_{13}=s_{13}^\magic$ of eq.~(\ref{eq:s13mgcIH}) for Case~I
results in an unobservable $\meg$ and only $\tmmm$ can be observed in addition to
$\tmee$ for all values of $m_3$.

\subsection{Case of $|h_{ee}|\simeq 0$}
An alternative way of suppressing BR($\meee$) is via small
$|h_{ee}|$. This scenario has been discussed~\cite{Akeroyd:2006bb}
in the context of the Left-Right symmetric model
in which the light neutrino mass matrix $m_{ij}$
in the expression for $h_{ij}$ in eq.~(\ref{hij}) is replaced
by an arbitrary mass matrix for heavy Majorana neutrinos.
We point out that $|h_{ee}|\simeq 0$ is also a viable option in the more restricted
framework of the HTM\@. The fact that $|m_{ee}|=0$ is possible for normal hierarchy
is well known in studies of neutrinoless double beta decay%
~\cite{KlapdorKleingrothaus:2000gr}.
 However, this possibility seems to have been overlooked
in the context of the HTM and its predictions for LFV decays.
Neutrinoless double beta decay in the HTM was
studied in \cite{Schechter:1981bd,Mohapatra:1981pm,Petcov:2009zr}.

\begin{figure}[t]
\begin{center}
\includegraphics[origin=c, angle=-90, scale=0.3]{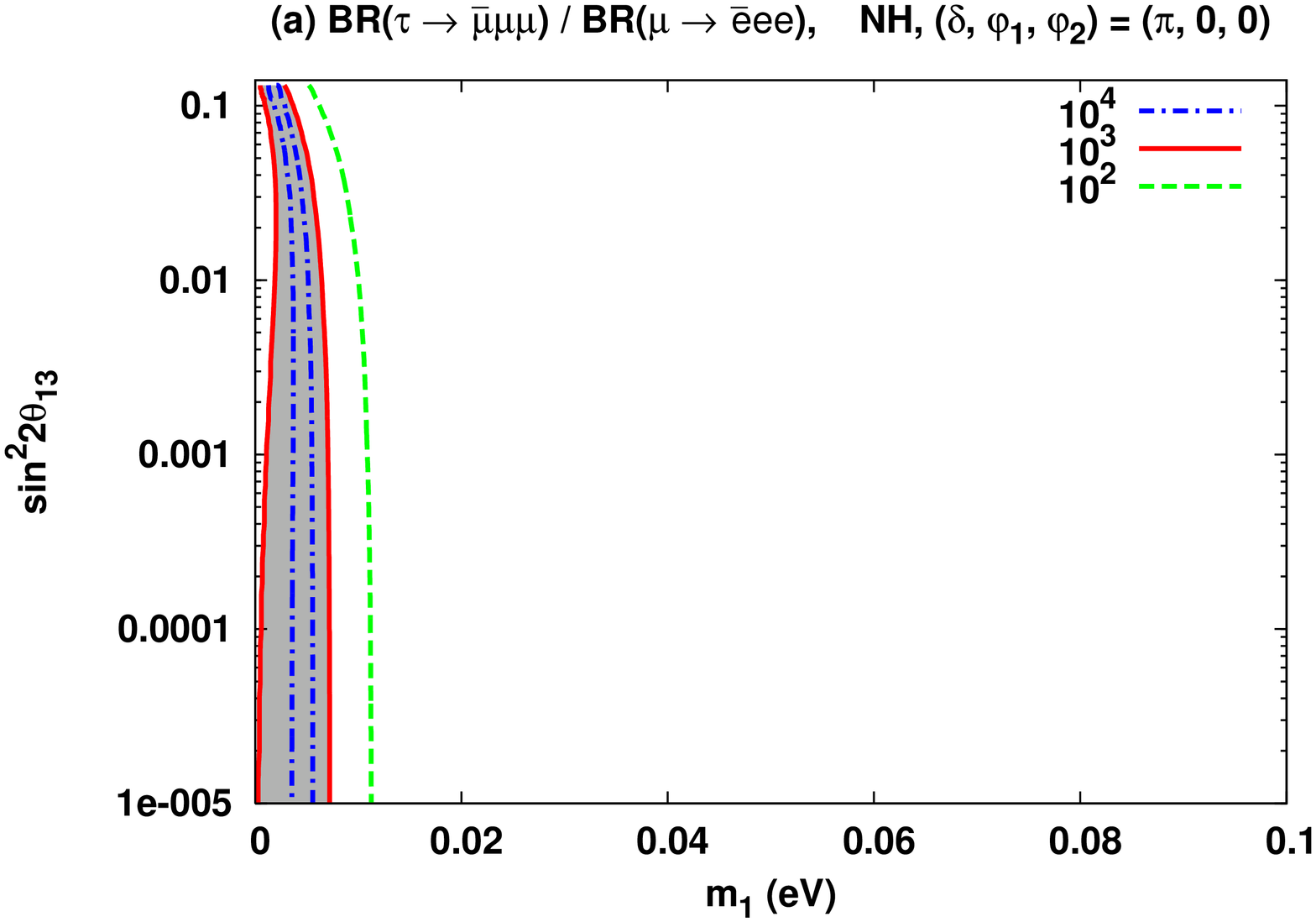}
\includegraphics[origin=c, angle=-90, scale=0.3]{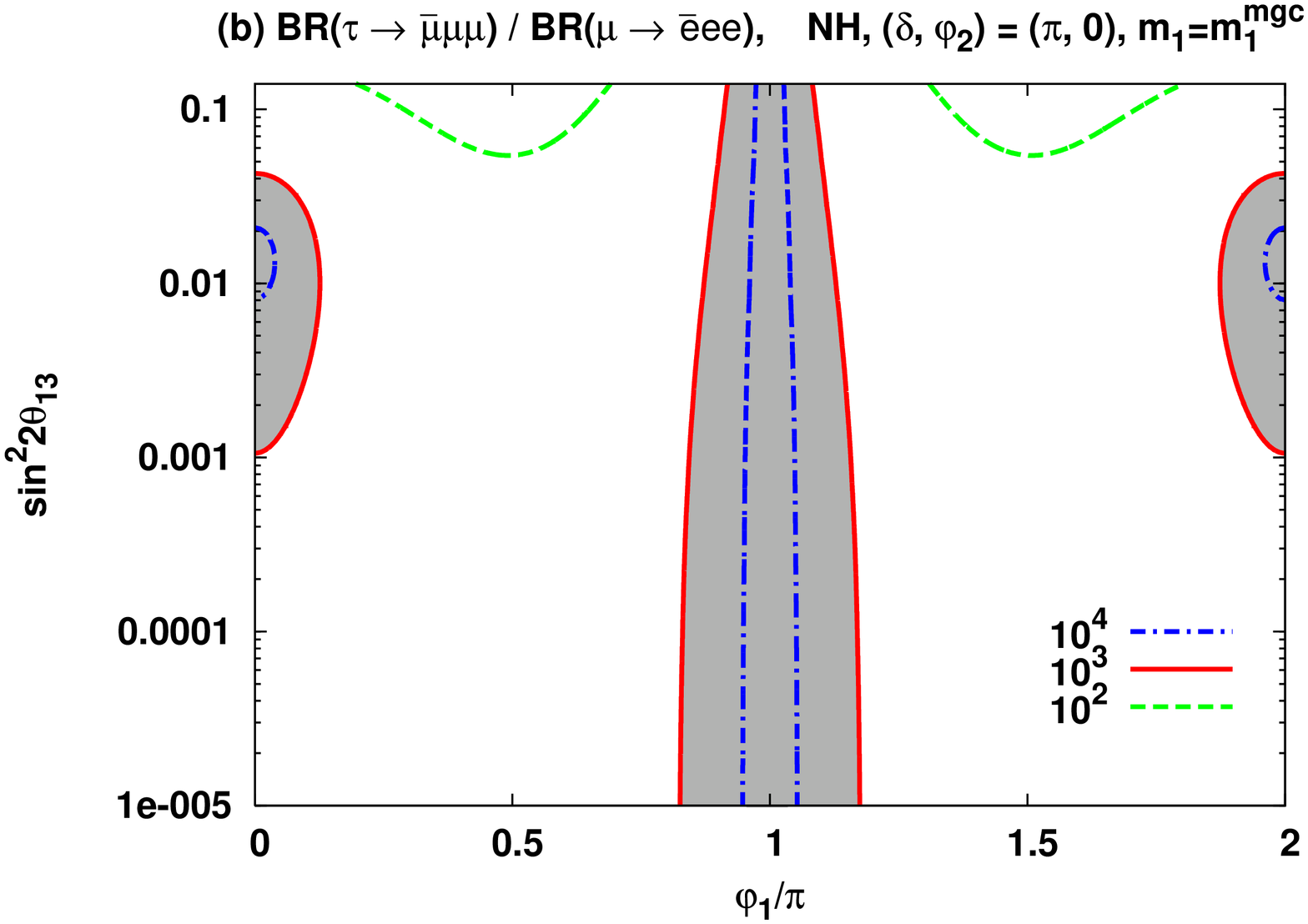}
\vspace*{-10mm}
\caption{
 Contours of \ratio{\tmmm}{\meee}
in the normal hierarchy.
 a) $m_1$-dependence
with $\varphi_1 = \varphi_1^\magic = \pi$
for $(\delta,\varphi_2) = (\pi, 0)$,
which corresponds to Case~III\@.
 b) $\varphi_1$-dependence
with $m_1=m_1^\magic$ for $(\delta,\varphi_2) = (\pi, 0)$,
where $m_1^\magic$ is a function of $\sin^2{2\theta_{13}}$.
A signal for $\tmmm$ is possible if the ratio $> 10^3$ (depicted by the shaded region).
}
\label{fig:tmmm_dp}
\end{center}
\end{figure}

\begin{figure}[t]
\begin{center}
\includegraphics[origin=c, angle=-90, scale=0.3]{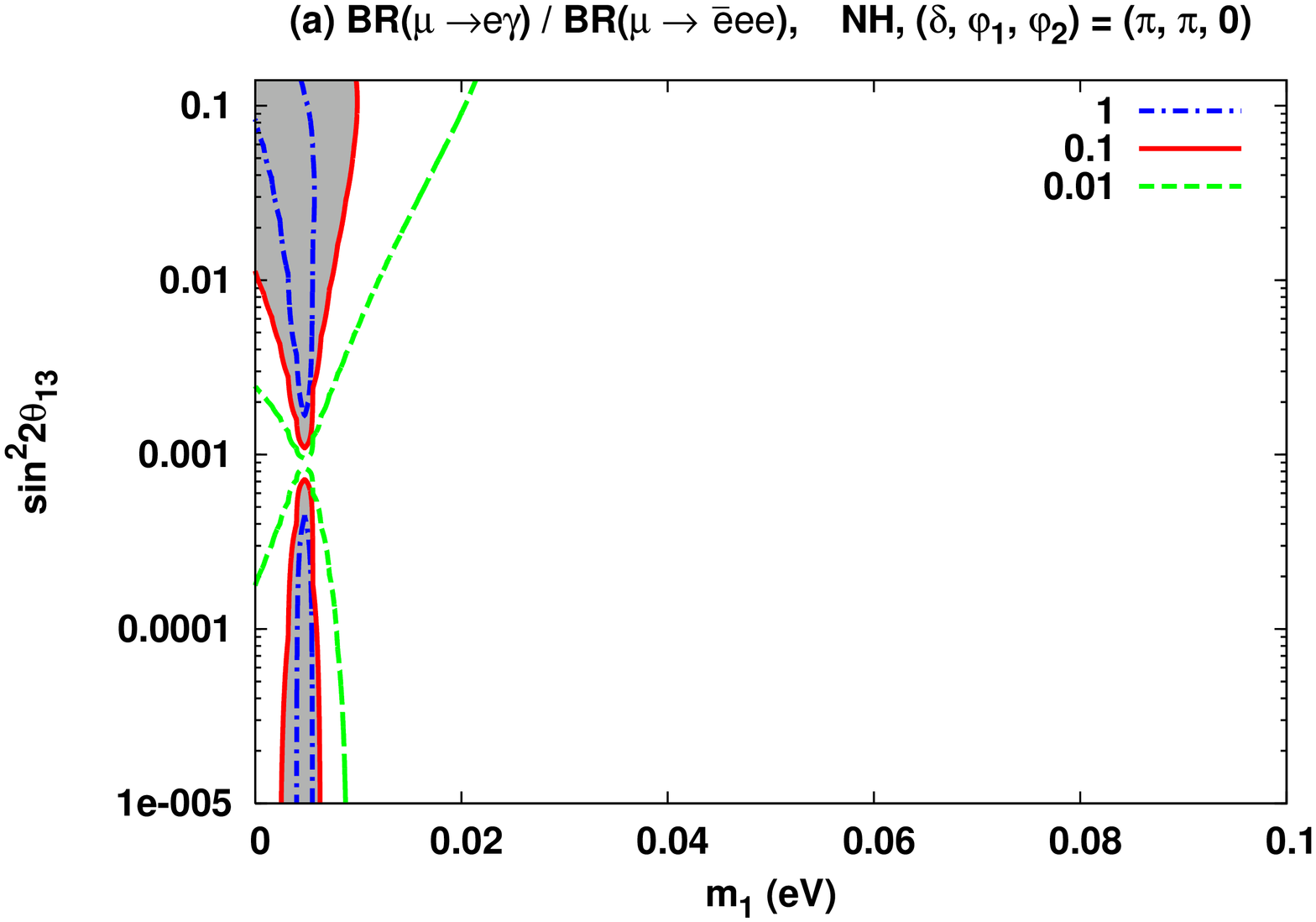}
\includegraphics[origin=c, angle=-90, scale=0.3]{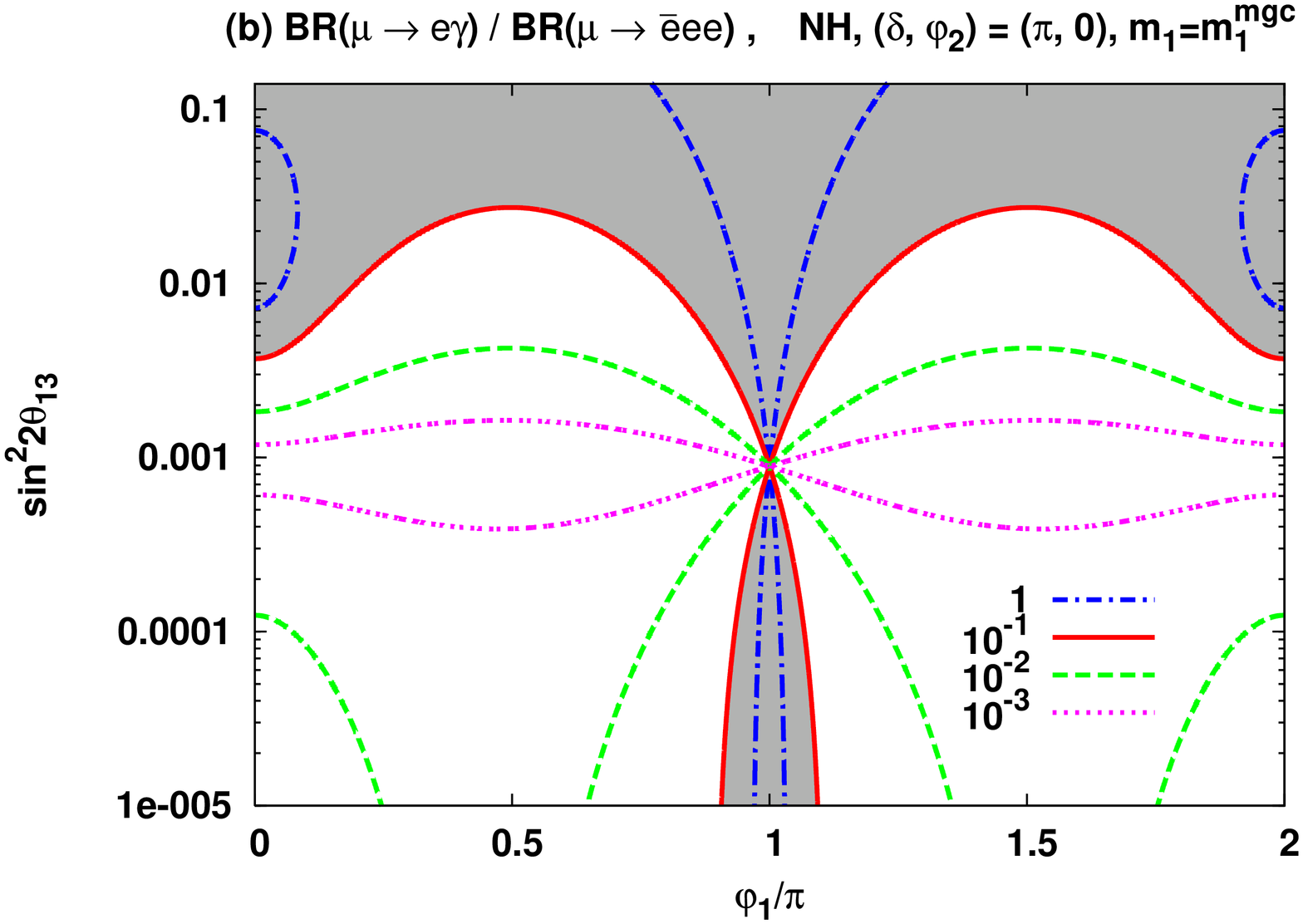}
\vspace*{-10mm}
\caption{
Same as Fig.~\ref{fig:tmmm_dp} but for $\meg$. A signal for $\mu\to e\gamma$ is 
possible if the ratio $> 0.1$ (depicted by the shaded region).
}
\label{fig:meg_dp}
\end{center}
\end{figure}

\subsubsection{Normal Hierarchy}
The conditions for $|h_{ee}|\simeq 0$ are different to those for $|h_{e\mu}|\simeq 0$,
and hence this possibility enlarges the parameter space for an observable signal
for $\tlll$ and $\mu\to e\gamma$. 
 Let us start with $m_1=0$ for simplicity.
 The complete elimination of $|h_{ee}|$ is possible, in principle,
with $\varphi_2-2\delta-\varphi_1=\pi$
at $s_{13}$ given by
\begin{eqnarray}
 s_{13}^2
 = \frac{ s_{12}^2\sqrt{\Delta m^2_{21}} }
        { s_{12}^2\sqrt{\Delta m^2_{21}} + \sqrt{\Delta m^2_{31}} }
 \simeq 0.05.
\label{eq:s13max}
\end{eqnarray}
 However, this value of $s_{13}^2$ violates the bound
$\sin^2{2\theta_{13}}< 0.14$ ($s_{13}^2 \lesssim 0.04$).
 On the other hand,
$|h_{ee}|=0$ is achieved
for given values of $s_{13}$ and $\varphi_2 - 2\delta$
by the magic values of $\varphi_1$ and $m_1$:
\begin{eqnarray}
\sin\varphi_1^\magic
&\equiv&
 - \frac{ \sqrt{ (m_1^\magic)^2 + \Delta m^2_{31} } }
        { \ s_{12}^2 \sqrt{ (m_1^\magic)^2 + \Delta m^2_{21} } \ }\,
   t_{13}^2 \sin(\varphi_2 - 2\delta), \ \ \
\cos\varphi_1^\magic \leq 0,
\label{eq:phi1mgc}
\\
(m_1^\magic)^2
&\equiv&
 \frac{1}{ \cos^2{2\theta_{12}}
           - 2 \left(
                s_{12}^4 + c_{12}^4 \cos{2(\varphi_2 - 2\delta)}
               \right) t_{13}^4
           + t_{13}^8 }
\nonumber\\
&&\hspace*{0mm}
\times
  \Biggl[ \
   s_{12}^4 \cos{2\theta_{12}} \Delta m^2_{21}
   + \left\{
      s_{12}^4 \Delta m^2_{21}
      + \left(
         s_{12}^4 + c_{12}^4 \cos{2(\varphi_2 - 2\delta)}
        \right) \Delta m^2_{31}
     \right\} t_{13}^4
\nonumber\\
&&\hspace*{40mm}
 {}- \Delta m^2_{31} t_{13}^8
 {}- 2 c_{12}^2 t_{13}^2 \cos(\varphi_2 - 2\delta)
     \sqrt{ A + B\,t_{13}^4 } \
  \Biggr],
\label{eq:m1mgc}
\\
A
&\equiv&
 \left(
  s_{12}^4 \Delta m^2_{21}
  + \cos{2\theta_{12}} \Delta m^2_{31} 
 \right)
 s_{12}^4 \Delta m^2_{21},\\
B
&\equiv&
 \Bigl\{
  ( s_{12}^4 - c_{12}^4 \sin^2(\varphi_2 - 2\delta) )
  \Delta m^2_{31}
  - s_{12}^4 \Delta m^2_{21}
 \Bigr\} \Delta m^2_{31},
\end{eqnarray}
where we define $t_{13}\equiv s_{13}/c_{13}$.
We have $|h_{ee}|=0$ at $s_{13}=0$
with $m_1^\magic \simeq 4.6\times 10^{-3}\,\eV$.
In contrast, $|h_{e\mu}|=0$ can only be obtained at $s_{13}=0$ in the 
unphysical limit of infinite $m_1$. Elimination of $h_{ee}$
is possible in Case~III and IV of $\varphi_1 = \pi$
but Case~I and II cannot realize $|h_{ee}|=0$
within the bound $\sin^2{2\theta_{13}} < 0.14$.

Fig.~\ref{fig:tmmm_dp}(a) shows
the ratio \ratio{\tmmm}{\meee} in the plane $[m_1$, $\sin^2{2\theta_{13}}]$
with $\varphi_1 = \varphi_1^\magic$
for $\delta = \pi$ and $\varphi_2 = 0$.
This case corresponds to Case~III
because these values of $\delta$ and $\varphi_2$
give $\varphi_1^\magic=\pi$.
The vertical funnel in the figure is caused by $|h_{ee}|\simeq 0$
around $m_1^\magic$ of eq.~(\ref{eq:m1mgc}).
One sees that the ratio $>10^{3}$ can be obtained
without fine-tuning of $m_1$ to $m_1^{\magic}$,
especially for very small $\sin^2{2\theta_{13}}$.
Since $|h_{ee}| = 0$ can be achieved for $s_{13}=0$ also,
there is no tuning of $\delta$ and $\varphi_2$,
which appear only with $s_{13}$ in $h_{ee}$.
Fig.~\ref{fig:tmmm_dp}(b) shows the $\varphi_1$-dependence
of the ratio 
with $m_1 = m_1^\magic$.  
Two isolated regions which give the ratio $>10^3$
exist around $\varphi_1=\pi$ and $0$.
The large ratio around $\varphi_1=\pi$ is caused by $|h_{ee}|\simeq 0$.
The other large ratio around $\varphi_1=0$ corresponds
to the suppression of $|h_{e\mu}|$ in Case~I
(see eq.~(\ref{eq:s13mgc-1})) and 
can also be seen in Fig.~\ref{fig:tmmm_hem0_n}(a)
at $m_1=m_1^\magic$.
It is evident that
sufficient suppression of $|h_{ee}|$ (and $|h_{e\mu}|$ also)
is achieved in a sizable range of $\varphi_1$.

The ratio \ratio{\meg}{\meee} is presented in Fig.~\ref{fig:meg_dp}(a)
in the plane $[m_1$, $\sin^2{2\theta_{13}}]$
with $\varphi_1 = \varphi_1^\magic$ for $\delta = \pi$ and $\varphi_2 = 0$.
It can be seen that the regions of large ratio roughly correspond to the 
regions of large ratio in Fig.~\ref{fig:tmmm_dp}(a).
In Fig.~\ref{fig:meg_dp}(b) the ratio becomes large for $\varphi_1 \simeq 0$ and $\pi$
because of $|h_{ee}|\simeq 0$ and $|h_{e\mu}|\simeq 0$ respectively
(as in Fig.~\ref{fig:tmmm_dp}(b)).
In both figures the ratio becomes very small for $\sin^2{2\theta_{13}} \simeq 10^{-3}$
because $|(hh^\dagger)_{e\mu}|=0$.

\begin{figure}[t]
\begin{center}
\includegraphics[origin=c, angle=-90, scale=0.3]{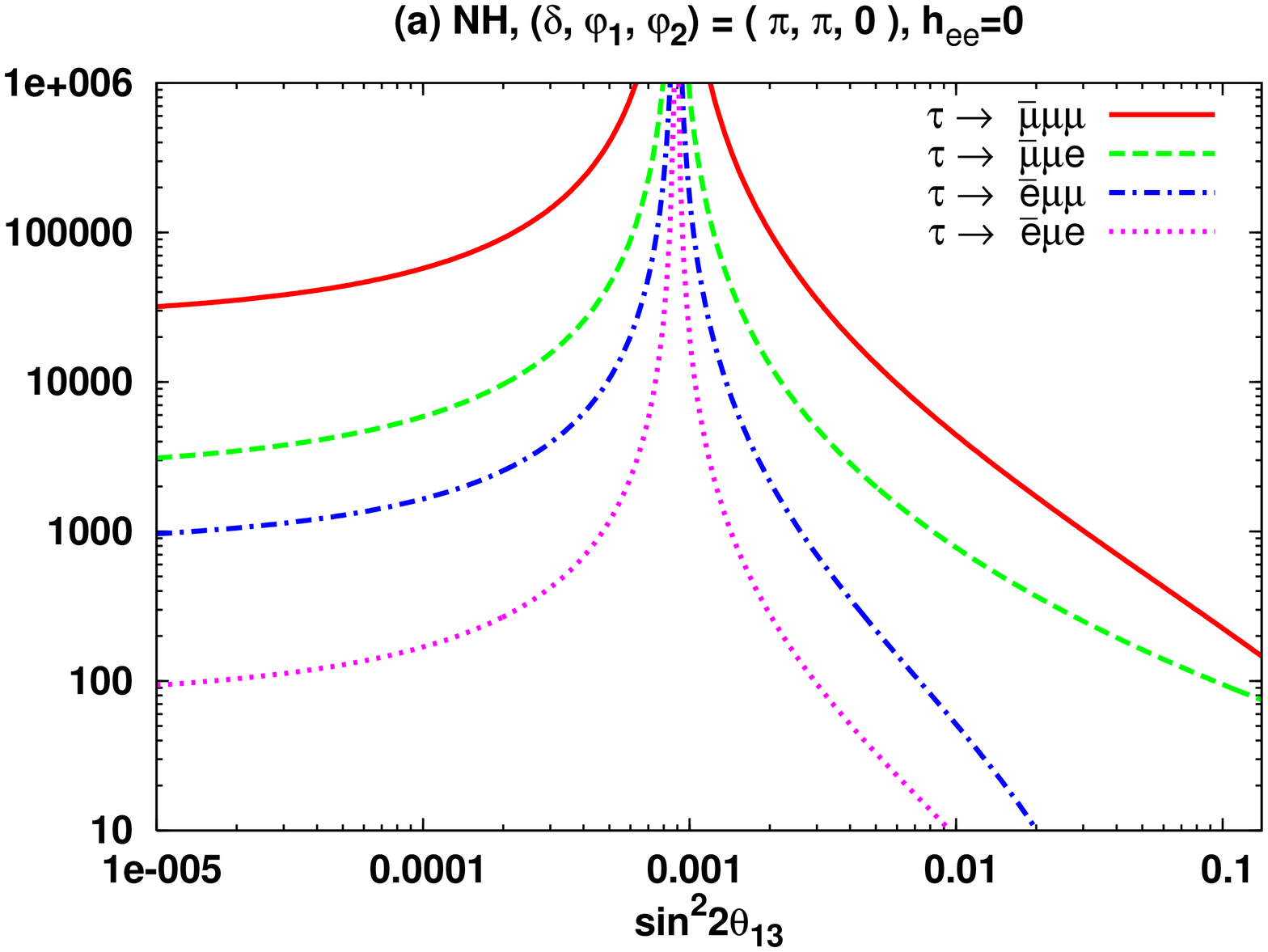}
\includegraphics[origin=c, angle=-90, scale=0.3]{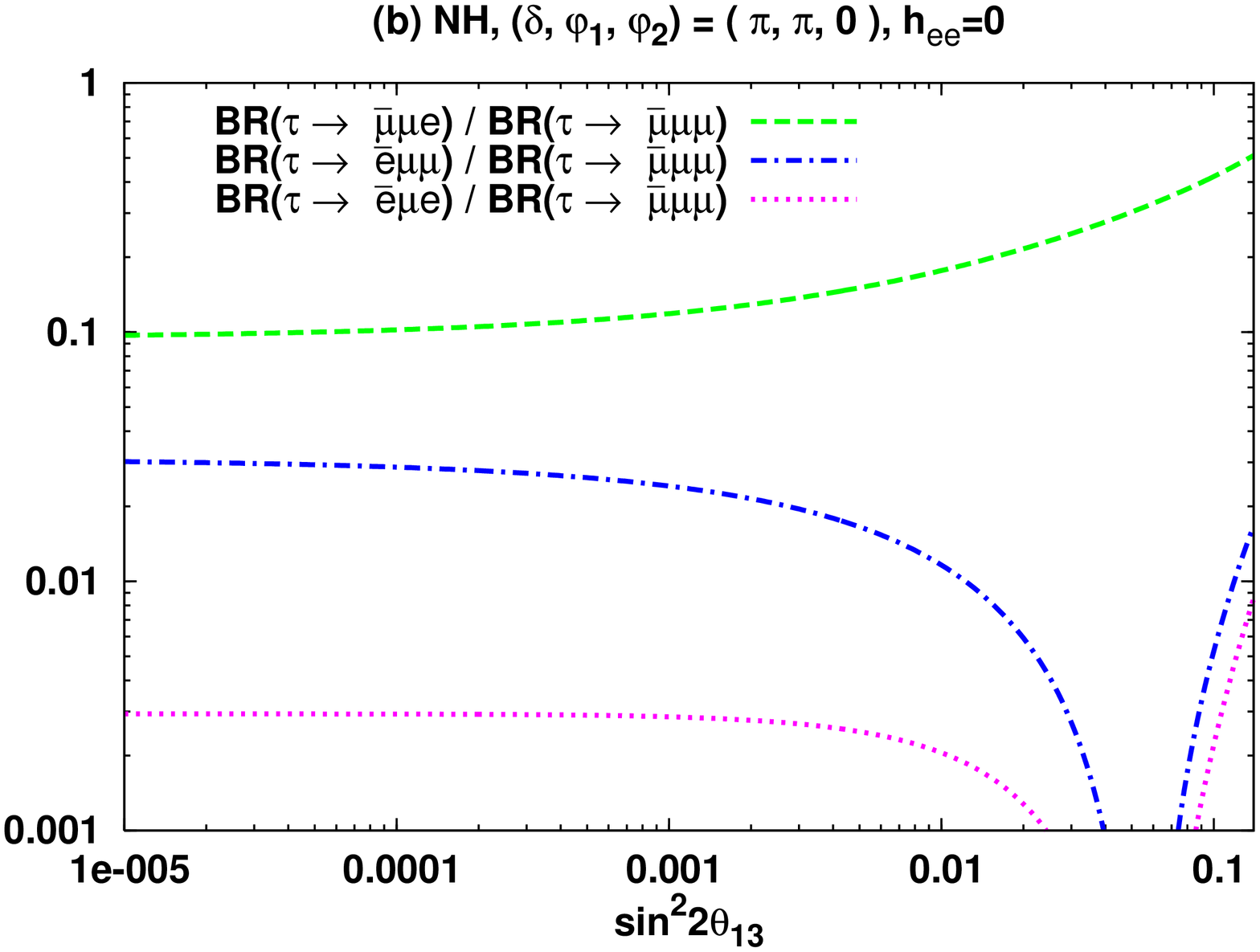}
\vspace*{-10mm}
\caption{
 $\sin^22\theta_{13}$-dependence of ratios of LFV decay branching ratios
for $|h_{ee}|=0$ with $m_1^\magic$ of eq.~(\ref{eq:m1mgc})
in Case~III in the normal hierarchy.
 $\delta = \pi$ is used.
 a) $\ratio{\tlll}{\meg}$;
 b) $\ratio{\tlll}{\tmmm}$.
}
\label{fig:ratio_dp_hee0}
\end{center}
\end{figure}

\begin{table}[t]
\begin{tabular}{c||c|c|c}
 & \multicolumn{3}{|c}{$(\delta, \varphi_1, \varphi_2) = (\pi, \pi, 0)$}\\
\cline{2-4}
\,
 &\, $\sin^2{2\theta_{13}} \lesssim 2\times 10^{-4}$\,
 &\, $2\times 10^{-4} \lesssim \sin^2{2\theta_{13}}
 \lesssim 2\times 10^{-3}$\,
 &\, $2\times 10^{-3}\lesssim \sin^2{2\theta_{13}}$\\
\hline\hline
NH
 & $\tmmm$
 & $\tmmm$
 & $\tmmm$\\[-3mm]
 & ($\tmme$)
 & $\tmme$
 & $\tmme$\\[-3mm]
 & $\meg$
 &
 & $\meg$
\end{tabular}
\caption{LFV decays which can be measured
for $|h_{ee}|=0$ in Case~III\@.
$\delta = \pi$ is used.
$\tmme$ with parentheses exists just below
the condition (\ref{cond:lll-2}).}
\label{tab:LFV_hee0}
\end{table}

Fig.~\ref{fig:ratio_dp_hee0} shows the $\theta_{13}$ dependence
of ratios of BRs for several LFV decays.
We take the magic value for $m_1$ of eq.~(\ref{eq:m1mgc})
in Case~III and the Dirac phase is taken as $\delta=\pi$.
 Fig.~~\ref{fig:ratio_dp_hee0}(a) shows $\ratio{\tlll}{\meg}$
and (b) presents $\ratio{\tlll}{\tmmm}$.
The solid, dashed, dash-dotted, and dotted lines show
results for $\tmmm$,
$\tau\to\overline{\mu}\mu e$,
$\tau\to\overline{e}\mu\mu$,
and $\tau\to\overline{e}\mu e$,
respectively. The decays $\tau\to\overline{e}ee$ and $\overline{\mu}ee$ are
forbidden via $|h_{ee}|= 0$.
In Fig.~\ref{fig:ratio_dp_hee0}(a),
the ratio for $\tmmm$ (the largest among $\tlll$)
becomes larger than $10^5$ for $\sin^2{2\theta_{13}}\simeq 10^{-3}$
because of $|(hh^\dagger)_{e\mu}|\simeq 0$
and $\meg$ cannot be measured in this region.
For $\sin^2{2\theta_{13}}$ near its maximum value, only the ratio for
$\tmmm$ stays $> 10^2$. Ratios for several $\tlll$ satisfy $>10^2$
for most values of $\sin^2{2\theta_{13}}$.
However, Fig.~\ref{fig:ratio_dp_hee0}(b) shows that
only $\tau\to \bar{\mu}\mu e$ can be measured
in addition to $\tmmm$ because the other decays do not satisfy
$\ratio{\tlll}{\tmmm}> 0.1$.
The decays $\temm$ and $\teme$ are vanishing
around $\sin^2{2\theta_{13}} = 7\times 10^{-2}$
because of $|h_{\tau e}| \simeq 0$.
 Table~\ref{tab:LFV_hee0} summarizes the
LFV decays which can be measured in this case.
We emphasize again that a signal of $\tmme$ is
impossible for the case where $\meee$
is suppressed by $|h_{e\mu}|\simeq 0$.

\begin{figure}[t]
\begin{center}
\includegraphics[origin=c, angle=-90, scale=0.3]{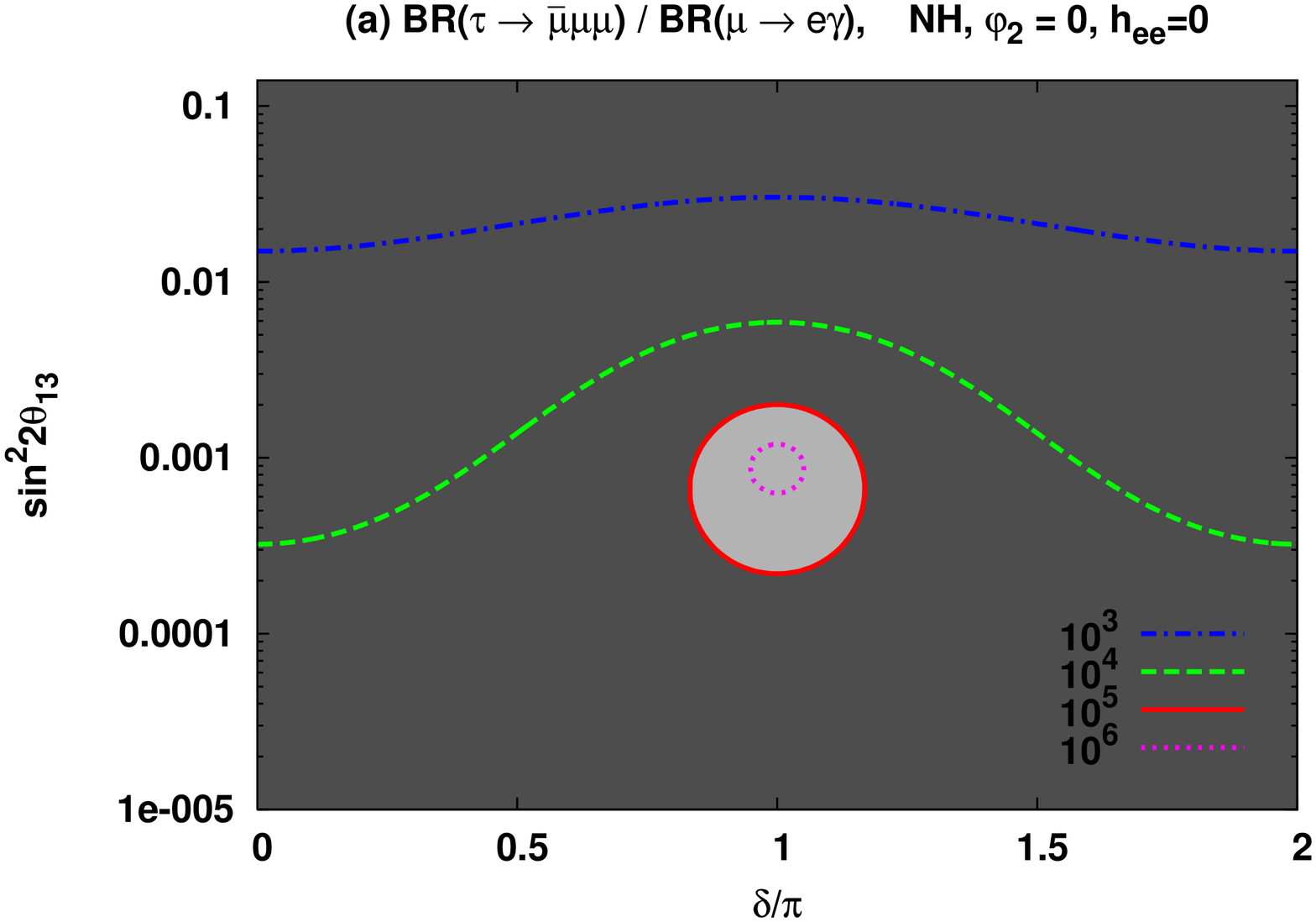}
\includegraphics[origin=c, angle=-90, scale=0.3]{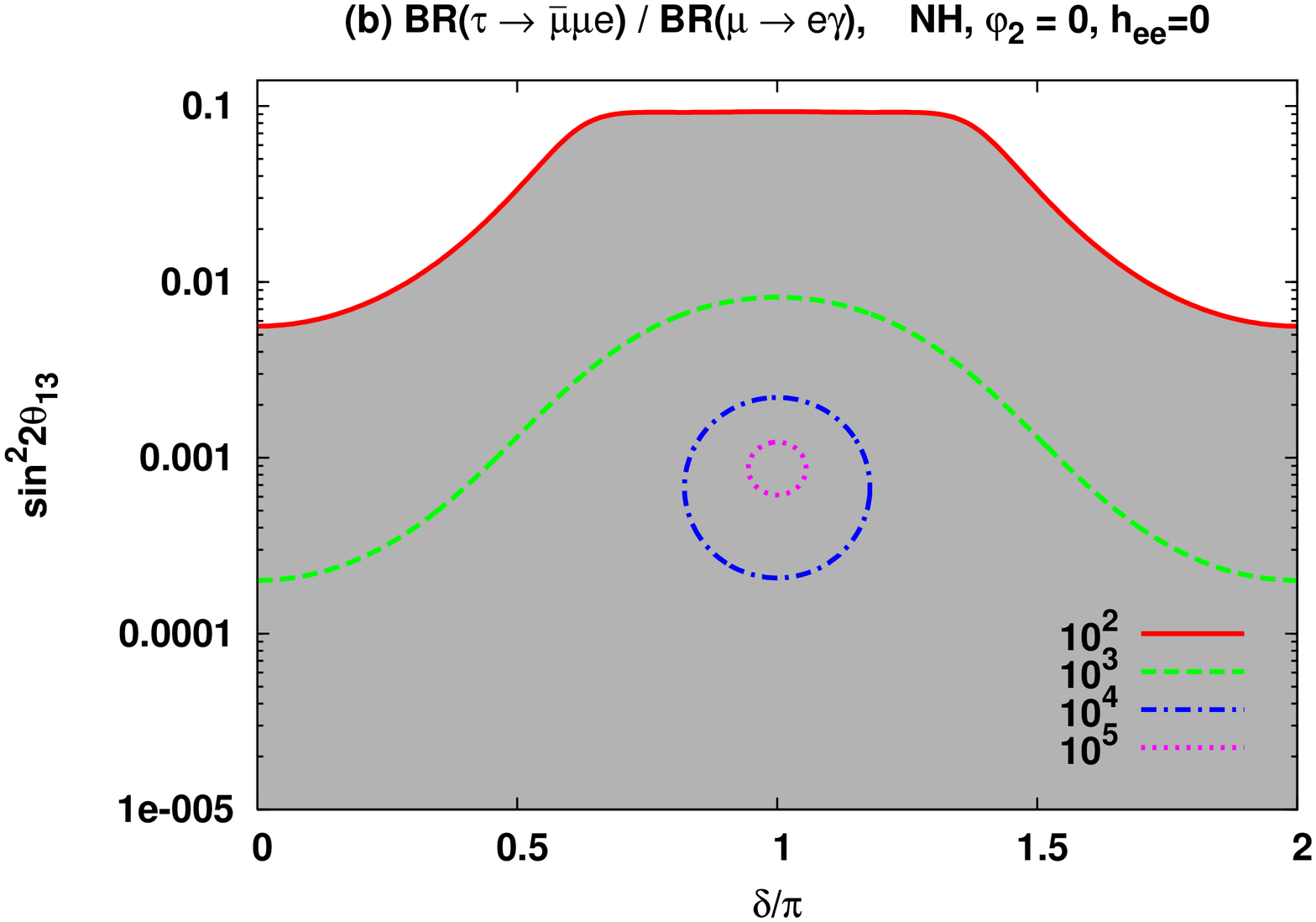}\\[-5mm]
\includegraphics[origin=c, angle=-90, scale=0.3]{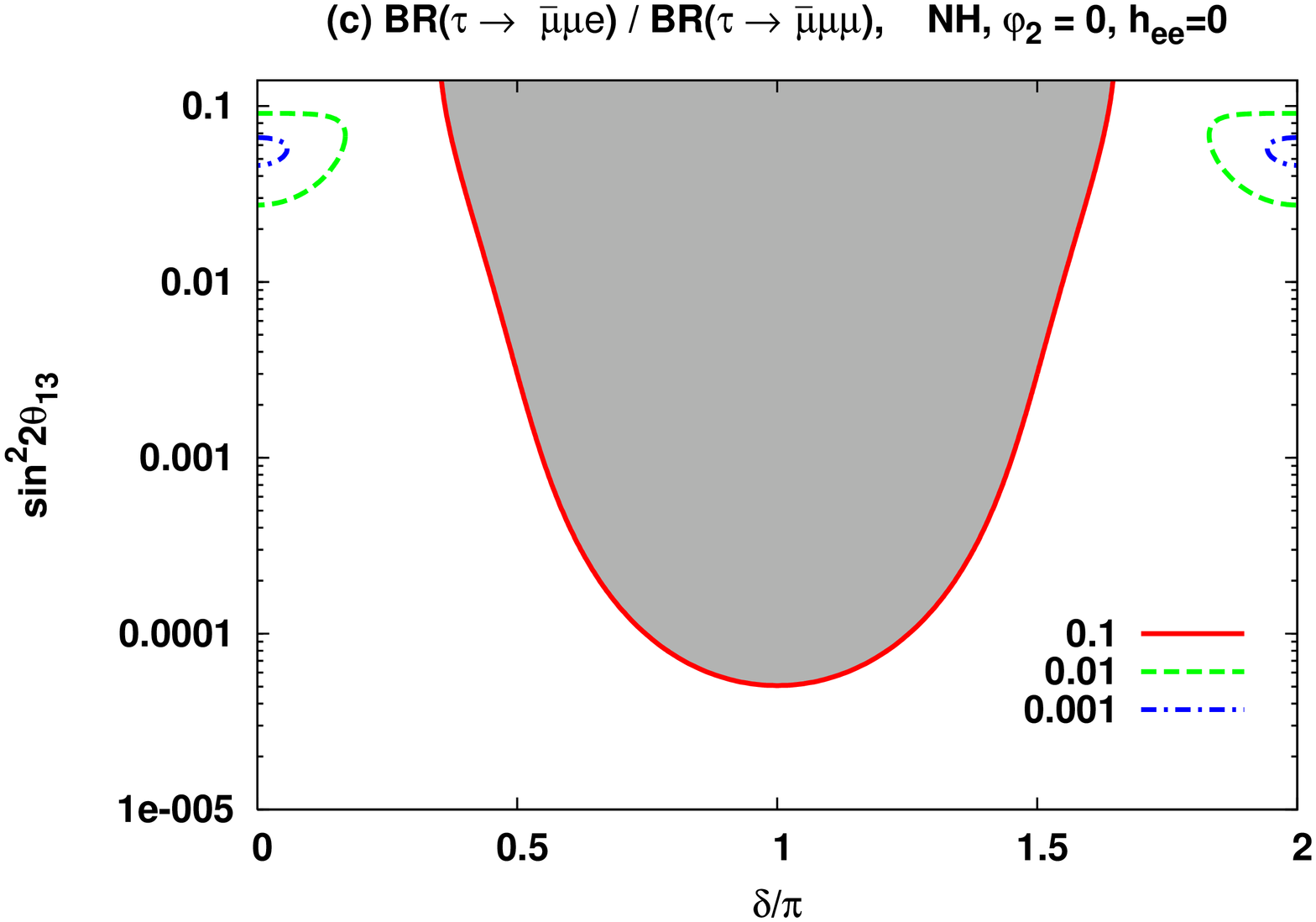}
\vspace*{-10mm}
\caption{
 Contours of ratios of BRs for $\varphi_2 = 0$ with $|h_{ee}|=0$.
(a) $\ratio{\tmmm}{\meg}$; the dark (light) shaded region signifies ratio $< 10^{5}$
($> 10^{5}$).
(b) $\ratio{\tmme}{\meg}$; the light shaded region signifies ratio $>10^{2}$.
(c) $\ratio{\tmme}{\tmmm}$; the light shaded region signifies ratio $>10^{-1}$.
}
\label{fig:meg}
\end{center}
\end{figure}

\vspace{5mm}
 Fig.~\ref{fig:meg}(a) shows
the ratio $\ratio{\tmmm}{\meg}$ 
in the plane $[\delta$, $\sin^2{2\theta_{13}}]$ for $\varphi_2=0$
with $|h_{ee}|=0$
which is achieved by $\varphi_1^\magic$ and $m_1^\magic$.
The ratio becomes very large
around $\delta = \pi$ and $\sin^2{2\theta_{13}}\simeq 1\times 10^{-3}$
because $|(hh^\dagger)_{e\mu}|\simeq 0$.
Except for this tiny region where the ratio is very large,
the ratio satisfies $\lesssim 10^{5}$ and $\gtrsim 10^{2}$
and therefore it is possible to observe a signal for
both $\mu\to e\gamma$ and $\tmmm$
in a very large part of the parameter space.
 As shown in Table~\ref{tab:LFV_hee0},
$\tmme$ can be measured in addition to $\tmmm$.
 The ratios for $\BR(\tmme)$ normalized by
$\BR(\meg)$ and $\BR(\tmmm)$ are
presented in Figs.~\ref{fig:meg}(b) and (c),
respectively.
 Note that a non-negligible rate for $\BR(\tmme)$ is
possible only if $\meee$ is suppressed by $|h_{ee}|\simeq 0$.
In fig.~\ref{fig:meg}(c) the ratio becomes very tiny
around $\delta=0$ and $\sin^2{2\theta_{13}}\simeq 7\times 10^{-2}$
because $|h_{e\mu}|\simeq 0$.
The ratio in Fig.~\ref{fig:meg}(b) exceeds $10^2$
in almost all of the plane $[\delta$, $\sin^2{2\theta_{13}}]$,
and the ratio in Fig.~\ref{fig:meg}(c) is larger than $0.1$
for $\sin^2{2\theta_{13}} \gtrsim 10^{-4}$ and for a wide range of values
of $\delta$ around $\delta=\pi$.
Even below the line for ratio $= 0.1$ in Fig.~\ref{fig:meg}(c),
the ratio is very close to 0.1 in most of the parameter space
as we have seen in Fig.~\ref{fig:ratio_dp_hee0}(b)
for $\delta=\pi$.
Therefore there are good prospects for observing
$\tmme$ in addition to $\tmmm$ in this scenario.

\subsubsection{Inverted Hierarchy}
The case of $|h_{ee}|\simeq 0$ cannot be realized
for the inverted hierarchy scenario under the constraints
on the neutrino oscillation parameters.
In the expression for $h_{ee}$ in eq.~(\ref{hij_expressions})
one sees that an exact cancellation between
the $m_1$ term and the $m_2$ term cannot be achieved
because $s_{12}^4/\cos{2\theta_{12}}\simeq 0.28 $ cannot be equal to
$m_1^2/\Delta m^2_{21}$, which is greater than
$|\Delta m^2_{31}|/\Delta m^2_{21}$ in the inverted hierarchy scenario.
Furthermore, $\theta_{13}$ is too small to cancel the remaining
difference between the $m_1$ term and the $m_2$ term.
In Fig.~\ref{fig:H++decay} (below), we will numerically
reconfirm that $|h_{ee}|\simeq 0$
is not possible in the inverted hierarchy scenario.

\subsection{Branching ratios of $H^{\pm\pm}\to l^\pm \l^\pm$ 
with observable LFV $\mu/\tau$ decay}

Finally, we discuss the impact of the observation of a LFV lepton decay
on the leptonic branching ratios of $H^{\pm\pm}$ in the HTM\@.
The LHC has sensitivity up to $m_{H^{\pm\pm}}\sim 1\TeV$ if 
$H^{\pm\pm}$ decays leptonically to $e^\pm e^\pm$, $e^\pm \mu^\pm$ or $\mu^\pm \mu^\pm$
with sizeable BRs~\cite{Gunion:1989in,Dion:1998pw,Akeroyd:2005gt,
Perez:2008ha,delAguila:2008cj}.
As can be seen from the Appendix, observation of a LFV decay
with $100\,\GeV < m_{H^{\pm\pm}}< 1000\,\GeV$ requires
$h_{ij}$ of the order $10^{-2}\,\text{-}\,10^{-3}$, which corresponds
to $1\,\eV \lesssim v_{\Delta} \lesssim 1000\,\eV$
(see eq.~(\ref{nu_mass})) with a naive bound
$\sqrt{|\Delta m^2_{31}|} \lesssim m_{ij} \lesssim 1\,\eV$.
For these values of $h_{ij}$ the decay width
for $\Hlilj$
(which is proportional to $v_{\Delta}^{-2}$)
is much larger than the decay width for the competing decay
$H^{\pm\pm}\to W^\pm W^\pm$ (which is proportional to $v^2_{\Delta}$)
and $H^{\pm\pm}\to H^\pm W^{\pm *}$ \cite{Akeroyd:2005gt,Chun:2003ej,Chakrabarti:1998qy}
(which is independent of $v_\Delta$ and potentially sizeable,
but is very suppressed for $m_{H^{\pm\pm}} > m_{H^\pm}$ if
the mass splitting is small, and absent if $m_{H^{\pm\pm}} < m_{H^\pm}$).
The decay branching ratios for $\Hlilj$ depend on just {\it one}
$h_{ij}$ coupling, while the LFV decays depend on the product of {\it two}
$h_{ij}$ couplings. Detailed studies of BR($\Hlilj$) have been performed
in \cite{Garayoa:2007fw,Akeroyd:2007zv,Kadastik:2007yd,Perez:2008ha}. 
We now impose the
condition for an observable LFV decay on the allowed regions of BR($\Hlilj$)
in the HTM\@. 
If $H^{\pm\pm}$ is observed at the LHC before a $\tau$ or $\mu$ LFV signal, {\sl and}
if BR$(H^{\pm\pm}\to l^\pm l^\pm)$ is consistent with the HTM prediction, then
measurements of BR$(H^{\pm\pm}\to l^\pm l^\pm)$ will 
determine whether a signal for LFV
violation is possible in the framework of the HTM\@.
Conversely, if a $\tau$ or $\mu$ LFV signal is observed first (which would also have an
interpretation in many models without $H^{\pm\pm}$), then the possible regions for 
BR$(H^{\pm\pm}\to l^\pm l^\pm)$ in the HTM would be constrained.

\begin{figure}[t]
\begin{center}
\includegraphics[origin=c, angle=-90, scale=0.35]{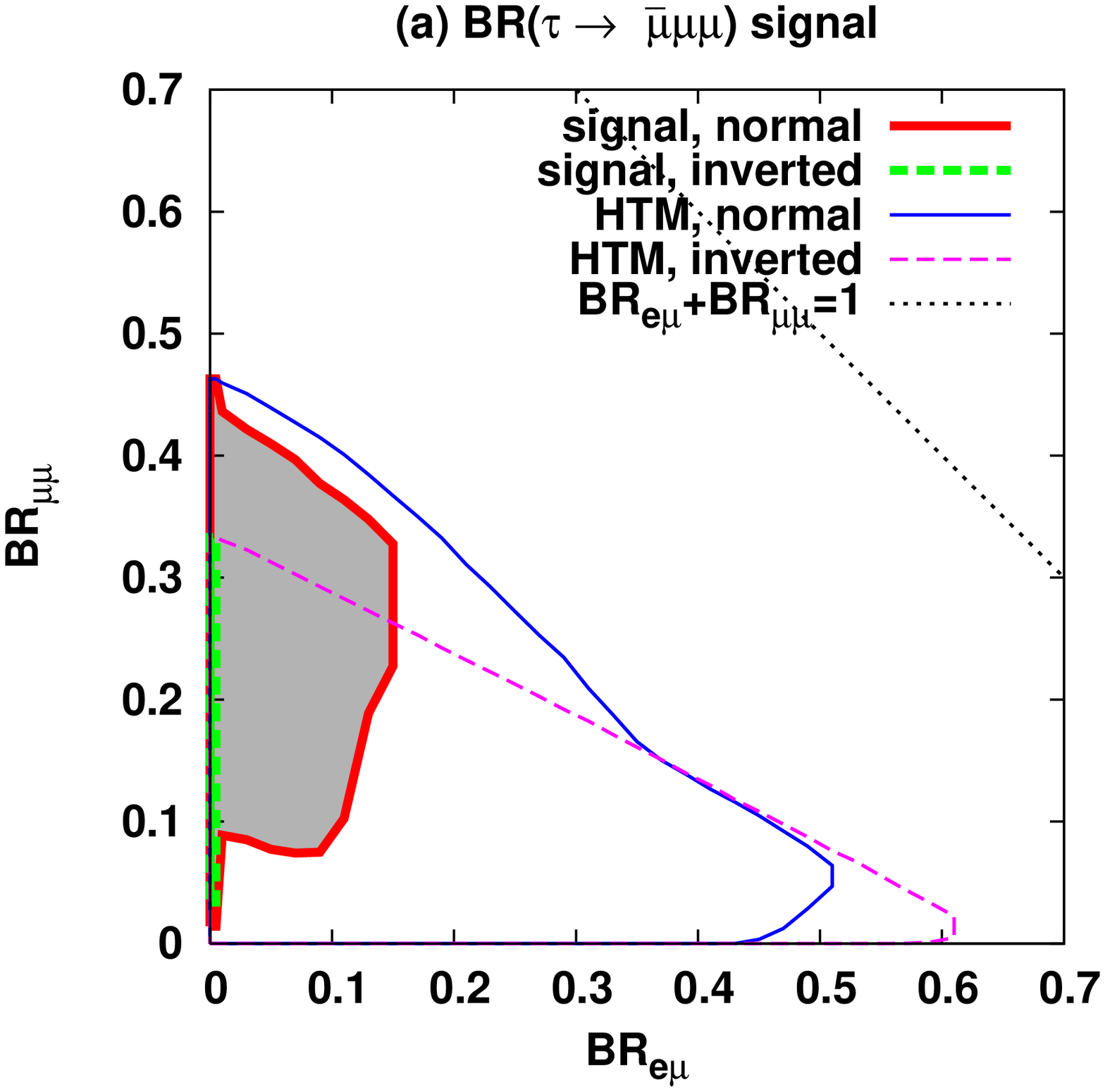}
\includegraphics[origin=c, angle=-90, scale=0.35]{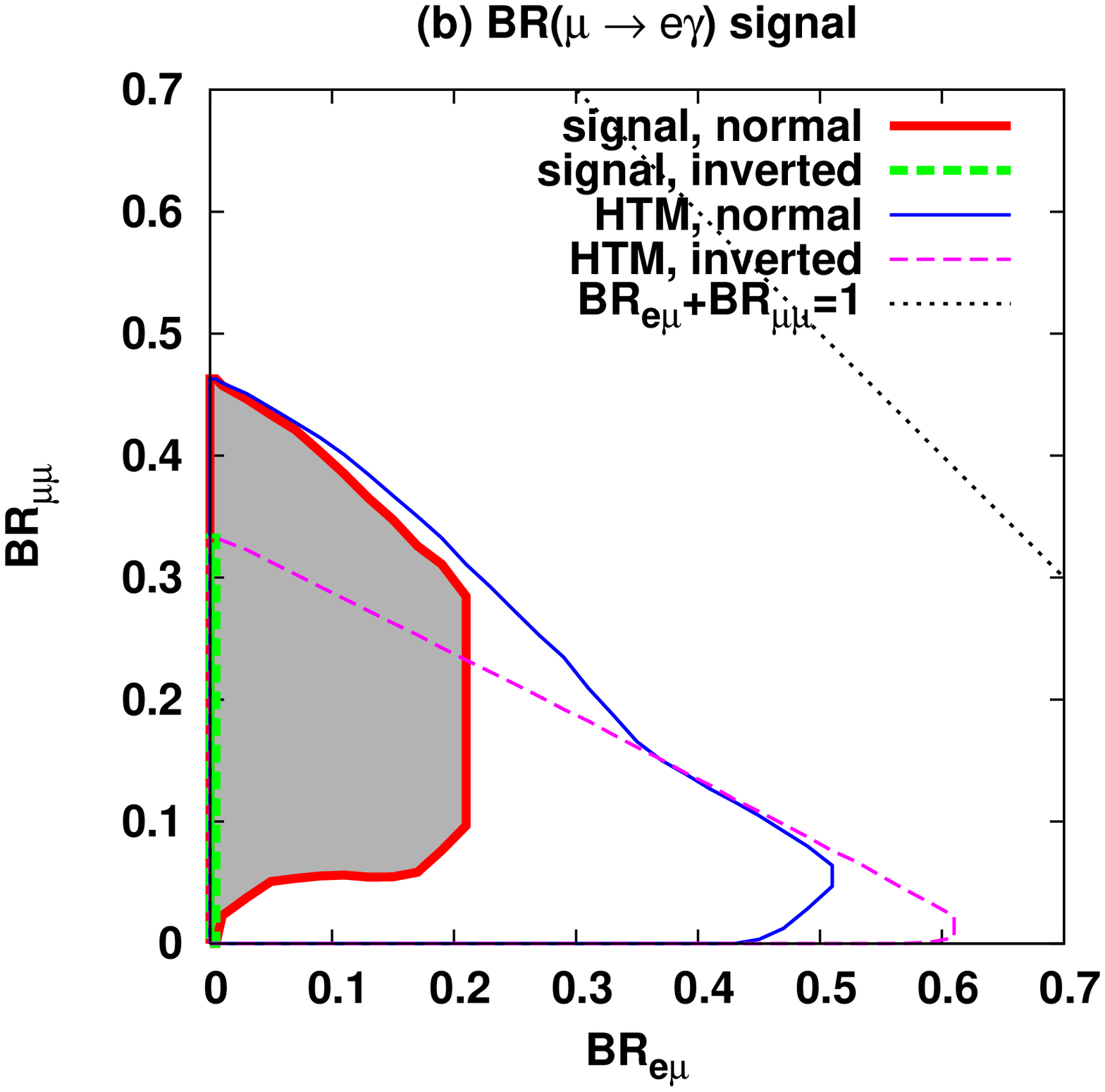}
\caption{
Region of allowed BR($H^{\pm\pm}\to e^\pm\mu^\pm$) and BR($H^{\pm\pm}\to \mu^\pm\mu^\pm$)
for normal and inverted hierarchy.
The range $0.94 \leq  \sin^2{2\theta_{23}} \leq 1$ is used:
a) Bold solid and bold dashed lines show the allowed BR regions after imposing the conditions for
an observable signal for $\tmmm$ (also depicted by the shaded region);
thin solid and thin dashed lines show the possible BR regions without imposing the 
conditions for an observable $\tmmm$.
b) Same as a) but for $\meg$.
}
\label{fig:H++decay}
\end{center}
\end{figure}

In Fig.~\ref{fig:H++decay}(a) the conditions $\ratio{\tmmm}{\meee} > 10^{3}$ 
and $\ratio{\tmmm}{\meg} > 10^{2}$
are imposed  on the plane of 
$\BR_{e\mu}$-$\BR_{\mu\mu}$, where
\begin{eqnarray}
 \BR_{ij}
 \equiv \BR(H^{\pm\pm}\to l_i^\pm l_j^\pm)
= \frac{S |m_{ij}|^2}
       {\sum_k m_k^2}.
\end{eqnarray}
In the figure $\sin^2{2\theta_{23}} > 0.94$%
~\cite{atm} was used.
The thin solid and thin dashed lines show
the possible regions in the HTM
for the normal and inverted hierarchies respectively, and correspond to
our previous result in \cite{Akeroyd:2007zv}.
The region above the dotted line
is unphysical because the sum of $\BR$s exceeds unity.
The bold solid and bold dashed lines
correspond to the boundaries of regions
in which an observable signal for $\tmmm$ is possible
for the normal and inverted hierarchies respectively.
 The HTM predicts $\BR_{e\mu} \simeq 0$
if $\meee$ is suppressed by $h_{e\mu}\simeq 0$. In the inverted hierarchy it is
clear that $\BR_{e\mu} \simeq 0$
is required for an observable signal for $\tmmm$,
i.e.\ $|h_{ee}|\sim 0$ cannot be realized.
In contrast, Fig.~\ref{fig:H++decay} shows that in the normal hierarchy scenario
$\BR_{e\mu}$ can reach $\sim 10\%$ because $\meee$ can be suppressed by $|h_{ee}|\simeq 0$.
However, the maximum value for $\BR_{e\mu}$ (for a given $\BR_{\mu\mu}$) is still considerably smaller than 
the maximum value for  $\BR_{e\mu}$ without imposing $\BR(\tmmm)/\BR(\meee) = 10^{3}$. 
The reason is because the condition $|h_{ee}|\simeq 0$
constrains $m_0$ which has a large effect on $\BR_{ij}$.
In contrast, $\BR_{ee}$ is not constrained so much
(beyond the allowed region in HTM)
because any value of $m_0$ is allowed if $\meee$ is suppressed by $|h_{e\mu}|\simeq 0$.
Small $\BR_{\mu\mu}$ is not preferred for $\tmmm$ signal with $|h_{ee}|=0$.
The result for $\meg$ is shown in Fig.~\ref{fig:H++decay}(b).
Inside the bold solid and bold dashed lines
one can have $\ratio{\meg}{\meee} > 0.1$ (and $\ratio{\meg}{\tmmm} < 10^5$ in order to
satisfy the bound on $\tmmm$)
for normal and inverted hierarchy respectively.
After imposing the condition for an observable BR($\mu \to e\gamma$)
 one can see from the figure that the maximum value of 
$\text{BR}_{e\mu}$ is about 20\%.
Observation of $\meg$ does not prefer
small $\BR_{\mu\mu}$ with $h_{ee}=0$ either.

\section{Conclusions}
The lepton flavour violating (LFV) decays 
$\tlll$ and $\meg$ 
were studied in the Higgs Triplet Model
(HTM). The stringent constraint from non-observation of
$\meee$ was discussed in detail, and corresponds
to distinct scenarios of $|h_{e\mu}|\simeq 0$ and $|h_{ee}|\simeq 0$.
The case of $|h_{e\mu}|= 0$ can be realized at a specific ("magic")
value of $\theta_{13}$ \cite{Chun:2003ej, Kakizaki:2003jk} which is obtained as a function
of the lightest neutrino mass ($m_0$) for both normal and inverted
neutrino mass hierarchies, and for each of the four
distinct cases for no CP violation from Majorana phases. 
The scenario of $|h_{ee}|= 0$ is only 
possible for normal neutrino mass hierarchy and requires
magic values of $\varphi_1$ and $m_1$, which are
functions of $\theta_{13}$ and $\varphi_2-2\delta$.
Observation of $\tlll$ at a proposed high luminosity $B$ factory 
requires its branching ratio to be $10^3$ larger than that of
$\meee$. It was shown that this can be realized for both 
scenarios $|h_{e\mu}|\simeq 0$ and $|h_{ee}|\simeq 0$ for a sizeable range of
$m_0$, $\theta_{13}$, and phases, and thus a signal for $\tlll$
does not require 
fine-tuning to the magic values of these parameters.
The pattern of branching ratios of $\tlll$ decays is
different for these two cases. A distinctive signal of the scenario
$|h_{ee}|\simeq 0$ would be the observation of $\tmme$, while
the observation of $\tmee$ indicates the scenario $|h_{e\mu}|\simeq 0$.
An observable signal for $\meg$ at the ongoing MEG experiment
requires a branching ratio  $>10^{-1}$ times that of $\mu \to \bar{e}ee$,
a condition which can be achieved in a wide region of 
the parameter space of the neutrino mass matrix. 
Finally, it was shown that detection of any of the above LFV decays 
in the HTM would significantly constrain the possible branching ratios
for the doubly charged Higgs boson to two leptons
($\Hlilj$). If $H^{\pm\pm}$ is observed at the Large Hadron Collider
then measurements of the branching ratios of $\Hlilj$ 
determine whether a signal of LFV in $\tau$ and/or $\mu$ decay
is possible or not in the HTM\@.

\section*{Acknowledgements}
A.G.A.\ is supported by the ``National Central University Plan to develop
First-class Universities and Top-level Research Centers''.
H.S.\ thanks IPMU where part of this work was done.
H.S.\ thanks K.~Tsumura for useful discussions.

\appendix
\section*{Appendix}

For comparison with the results of \cite{Chun:2003ej} 
we show explicit expressions for $h_{ij}$ and 
$(hh^\dagger)_{ij}$ using their approximations. 
The following seven distinct scenarios were considered
in \cite{Chun:2003ej}:

\begin{tabular}{rll}
HI
 &: $m_1 \ll m_2 < m_3$,
 & \ $\varphi_1 = \varphi_2 = 0$\\
IN1
 &: $m_3 \ll m_1 < m_2$,
 & \ $\varphi_1 = \varphi_2 = 0$\\
IN2
 &: $m_3 \ll m_1 < m_2$,
 & \ $\varphi_1 = \pi, \ \varphi_2 = 0$\\
DG1
 &: $\sqrt{|\Delta m^2_{31}|} \ll m_i$,
 & \ $\varphi_1 = \varphi_2 = 0$\\
DG2
 &: $\sqrt{|\Delta m^2_{31}|} \ll m_i$,
 & \ $\varphi_1 = 0, \ \varphi_2 = \pi$\\
DG3
 &: $\sqrt{|\Delta m^2_{31}|} \ll m_i$,
 & \ $\varphi_1 = \pi, \ \varphi_2 = 0$\\
DG4
 &: $\sqrt{|\Delta m^2_{31}|} \ll m_i$,
 & \ $\varphi_1 = \varphi_2 = \pi$
\end{tabular}
\vspace*{1mm}\\
\noindent
Note that
\begin{eqnarray}
&&
 \sqrt{2} v_\Delta h_{ij}
 = [ V_{MNS}
     \text{diag} (m_1, m_2 e^{i\varphi_1}, m_3 e^{i\varphi_2})
     V_{MNS}^T ]_{ij},\\
&&
 2v_\Delta^2 [hh^\dagger]_{ij}
 = m_1^2 \delta_{ij}
  + [ V_{MNS}
     \text{diag} (0, \Delta m^2_{21}, \Delta m^2_{31})
     V_{MNS}^\dagger ]_{ij}.
\end{eqnarray}

We find good agreement with the analytical results of
\cite{Chun:2003ej} except for a few cases which are
highlighted below. We note that $f_{ij}$ in \cite{Chun:2003ej} (Yukawa coupling for the Higgs triplet)
is related to $h_{ij}$ by $f_{ij}=\sqrt{2}\,h_{ij}$.
For the bound on $|(h h^\dagger)_{e\mu}|$ from
$\BR(\mu\to e\gamma)$ we find:\\
\begin{tabular}{ccc}
$\BR(\mu\to e\gamma) < 1.2\times 10^{-11}$
 & \ $\Rightarrow$ \
 & $|(h h^\dagger)_{e\mu}|
   < 5.2\times 10^{-5}
     \left( \frac{m_{H^{\pm\pm}}}{200\GeV} \right)^2$
\end{tabular}
\vspace*{1mm}\\
\noindent
This result disagrees with the bound for $(f f^\dagger)_{e\mu}$
given in Table~1 in \cite{Chun:2003ej}.
If we neglect the $H^\pm$ contribution to 
$\mu \to e\gamma$ in eq.(\ref{eq:ar}) we find better agreement  
with the bound on $(ff^\dagger)_{e\mu}$ in \cite{Chun:2003ej}.
For specific $\tlll$ decays we find:\\
\begin{tabular}{ccc}
$\BR(\tau\to \overline{\mu} e\mu) < 3.1\times 10^{-7}$
 & \ $\Rightarrow$ \
 & $|h_{e\mu} h_{\mu\tau}|
   < 8.9\times 10^{-4}
     \left( \frac{m_{H^{\pm\pm}}}{200\GeV} \right)^2$\\
$\BR(\tmmm) < 3.8\times 10^{-7}$
 & \ $\Rightarrow$ \
 & $|h_{\mu\mu} h_{\mu\tau}|
   < 1.4\times 10^{-3}
     \left( \frac{m_{H^{\pm\pm}}}{200\GeV} \right)^2$
\end{tabular}
\vspace*{1mm}\\
\noindent
The bound on $|h_{e\mu} h_{\mu\tau}|$ agrees with the result of
\cite{Chun:2003ej} but the bound on $|h_{\mu\mu} h_{\mu\tau}|$ disagrees.
It seems that a factor of 2 has been used for $\BR(\tmmm)$ in \cite{Chun:2003ej}
instead of 1/2 due to identical particles (two $\mu$'s)
in the final state.

For the case of $h_{e\mu}=0$ we find the following ratios of BRs:

\begin{tabular}{rl}
(HI) \
 & $\BR(\tau\to \overline{\mu}\mu\mu)
    :\BR(\mu\to e\gamma)
    = 1
      :4.6\times 10^{-2}
        \frac{\Delta m^2_{21}}{|\Delta m^2_{31}|}
        \sin^2{2\theta_{12}}$\\
(IN1) \
 & $\BR(\tau\to \overline{\mu}ee)
    :\BR(\tau\to \overline{\mu}\mu\mu)
    :\BR(\mu\to e\gamma)$\\
 &\hspace*{51mm}
    $= 1
      :0.25
      :2.9\times 10^{-3}
        \left( \frac{\Delta m^2_{21}}{|\Delta m^2_{31}|} \right)^2
        \sin^2{2\theta_{12}}$\\
(DG1) \
 & $\BR(\tau\to \overline{\mu}ee)
    :\BR(\tau\to \overline{\mu}\mu\mu)
    :\BR(\mu\to e\gamma)$\\
 &\hspace*{51mm}
   $= 1
     :1
     :2.9\times 10^{-3}
       \left( \frac{|\Delta m^2_{31}|}{m_1^2} \right)^2
       \left( \frac{\Delta m^2_{21}}{|\Delta m^2_{31}|} \right)^2
       \sin^2{2\theta_{12}}$\\
(DG2) \
 & $\BR(\tau\to \overline{\mu}ee)
    :\BR(\mu\to e\gamma)
    = 1
      :2.9\times 10^{-3}
        \left( \frac{|\Delta m^2_{31}|}{m_1^2} \right)^2
        \left( \frac{\Delta m^2_{21}}{|\Delta m^2_{31}|} \right)^2
        \sin^2{2\theta_{12}}$
\end{tabular}
\vspace*{1mm}\\
\noindent
A factor $8.6\times 10^{-3}$ for HI appears in \cite{Chun:2003ej}
instead of our $4.6\times 10^{-2}$.
This disagreement seems to be caused by the aforementioned differences
in $\BR(\tmmm)$ (an extra factor of 4 compared to ours)
and in $\BR(\mu\to e\gamma)$ (an extra factor of $(8/9)^2$ compared to ours).
The ratio $\BR(\meg)/\BR(\tmee)$ is the same in DG1 and DG2,
although there is no $\meg$ in DG1 in \cite{Chun:2003ej}.

In Tables \ref{tab:hij} and \ref{tab:hhij}
we present expressions for $h_{ij}$ and $[hh^\dagger]_{ij}$.
The Dirac phase is taken to be zero ($\delta=0$).
Explicit $\delta$ dependence is not shown but enters through $s_{13}$.
We find agreement with the results of \cite{Chun:2003ej} except for 
the following cases: $h_{\mu\mu}$ and $h_{\tau\tau}$ in IN2;
$h_{e\mu}$ and $h_{e\tau}$ in DG2; $[hh^\dagger]_{e\mu}$ 
and $[hh^\dagger]_{e\tau}$ in IN.

\begin{sidewaystable}
\begin{flushleft}
\begin{tabular}{|c||c|c|c}
 & $\sqrt{2} v_\Delta h_{ee}$
 & $\sqrt{2} v_\Delta h_{\mu\mu}$
 & $\sqrt{2} v_\Delta h_{\tau\tau}$\\
\hline\hline
\text{HI}
 & $\sqrt{|\Delta m^2_{31}|}
   \left(
    s_{13}^2 + \sqrt{\frac{\Delta m^2_{21}}{|\Delta m^2_{31}|}} s_{12}^2
   \right)$
 & $\frac{\sqrt{|\Delta m^2_{31}|}}{2}$
 & $\frac{\sqrt{|\Delta m^2_{31}|}}{2}$\\
%
%
\hline
\text{IN1}
 & $\sqrt{|\Delta m^2_{31}|}$
 & $\frac{\sqrt{|\Delta m^2_{31}|}}{2}$
 & $\frac{\sqrt{|\Delta m^2_{31}|}}{2}$\\
\hline
\text{IN2}
 & $\sqrt{|\Delta m^2_{31}|} \cos{2\theta_{12}}$
 & $-\frac{\sqrt{|\Delta m^2_{31}|}}{2}
   \left(
    \cos{2\theta_{12}} - 2 s_{13} \sin{2\theta_{12}}
   \right)$
 & $-\frac{\sqrt{|\Delta m^2_{31}|}}{2}
   \left(
    \cos{2\theta_{12}} + 2 s_{13} \sin{2\theta_{12}}
   \right)$\\
\hline
\text{DG1}
 & $m_1$
 & $m_1$
 & $m_1$\\
\hline
\text{DG2}
 & $m_1$
 & $m_1 
   \left(
    s_{13}^2 + \cos{2\theta_{23}}
    - \frac{1}{\,4\,} \frac{\Delta m^2_{31}}{m_1^2}
   \right)$
 & $m_1 
   \left(
    s_{13}^2 - \cos{2\theta_{23}}
    - \frac{1}{\,4\,} \frac{\Delta m^2_{31}}{m_1^2}
   \right)$\\
\hline
\text{DG3}
 & $m_1 \cos{2\theta_{12}}$
 & $m_1 \left( s_{12}^2 + s_{13} \sin{2\theta_{12}} \right)$
 & $m_1 \left( s_{12}^2 - s_{13} \sin{2\theta_{12}} \right)$\\
\hline
\text{DG4}
 & $m_1 \cos{2\theta_{12}}$
 & $-m_1 \left( c_{12}^2 - s_{13}\sin{2\theta_{12}} \right)$
 & $-m_1 \left( c_{12}^2 + s_{13}\sin{2\theta_{12}} \right)$\\
\end{tabular}
\end{flushleft}
\begin{flushright}
\begin{tabular}{c|c|c|}
$\sqrt{2} v_\Delta h_{e\mu}$
 & $\sqrt{2} v_\Delta h_{e\tau}$
 & $\sqrt{2} v_\Delta h_{\mu\tau}$\\
\hline\hline
$\frac{\sqrt{|\Delta m^2_{31}|}}{\sqrt{2}}
   ( s_{13}
     + \frac{1}{\,2\,} \sqrt{\frac{\Delta m^2_{21}}{|\Delta m^2_{31}|}}
       \sin{2\theta_{12}} )$
 & $\frac{\sqrt{|\Delta m^2_{31}|}}{\sqrt{2}}
   ( s_{13}
     - \frac{1}{\,2\,} \sqrt{\frac{\Delta m^2_{21}}{|\Delta m^2_{31}|}}
       \sin{2\theta_{12}} )$
 & $\frac{\sqrt{|\Delta m^2_{31}|}}{2}$\\
%
\hline
$-\frac{\sqrt{|\Delta m^2_{31}|}}{\sqrt{2}}
   ( s_{13}
     - \frac{1}{\,4\,}
       \frac{\Delta m^2_{21}}{|\Delta m^2_{31}|} \sin{2\theta_{12}} )$
 & $-\frac{\sqrt{|\Delta m^2_{31}|}}{\sqrt{2}}
   ( s_{13}
     + \frac{1}{\,4\,}
       \frac{\Delta m^2_{21}}{|\Delta m^2_{31}|} \sin{2\theta_{12}} )$
 & $-\frac{\sqrt{|\Delta m^2_{31}|}}{2}$\\
\hline
$-\frac{\sqrt{|\Delta m^2_{31}|}}{\sqrt{2}} \sin{2\theta_{12}}$
 & $-\frac{\sqrt{|\Delta m^2_{31}|}}{\sqrt{2}} \sin{2\theta_{12}}$
 & $\frac{\sqrt{|\Delta m^2_{31}|}}{2} \cos{2\theta_{12}}$\\
\hline
$\frac{\Delta m^2_{31}}{2\sqrt{2} m_1}
   ( s_{13}
     + \frac{1}{\,2\,}
       \frac{\Delta m^2_{21}}{\Delta m^2_{31}} \sin{2\theta_{12}})$
 & $\frac{\Delta m^2_{31}}{2\sqrt{2} m_1}
   ( s_{13}
     - \frac{1}{\,2\,}
       \frac{\Delta m^2_{21}}{\Delta m^2_{31}} \sin{2\theta_{12}})$
 & $\frac{\Delta m^2_{31}}{4 m_1}$\\
\hline
$-\sqrt{2} m_1
   ( s_{13}
     - \frac{1}{\,8\,}
       \frac{\Delta m^2_{31}}{m_1^2}
       \frac{\Delta m^2_{21}}{\Delta m^2_{31}} \sin{2\theta_{12}} )$
 & $-\sqrt{2} m_1
   ( s_{13}
     + \frac{1}{\,8\,}
       \frac{\Delta m^2_{31}}{m_1^2}
       \frac{\Delta m^2_{21}}{\Delta m^2_{31}} \sin{2\theta_{12}} )$
 & $-m_1$\\
\hline
$-\frac{m_1}{\sqrt{2}} (\sin{2\theta_{12}} - 2 s_{13}s_{12}^2)$
 & $\frac{m_1}{\sqrt{2}} (\sin{2\theta_{12}} + 2 s_{13}s_{12}^2)$
 & $m_1 c_{12}^2$\\
\hline
$-\frac{m_1}{\sqrt{2}} (\sin{2\theta_{12}} + 2s_{13}c_{12}^2)$
 & $\frac{m_1}{\sqrt{2}} (\sin{2\theta_{12}} - 2s_{13}c_{12}^2)$
 & $-m_1 s_{12}^2$
\end{tabular}
\end{flushright}
\caption{Approximate forms of $\sqrt{2} v_\Delta h_{ij}$.}
\label{tab:hij}
\end{sidewaystable}

\begin{sidewaystable}
\begin{center}
\begin{tabular}{|c||c|c|c|c|c|c|}
 & $2v_\Delta^2 [hh^\dagger]_{ee}$
 & $2v_\Delta^2 [hh^\dagger]_{\mu\mu}$
 & $2v_\Delta^2 [hh^\dagger]_{\tau\tau}$
 & $2v_\Delta^2 [hh^\dagger]_{e\mu}$
 & $2v_\Delta^2 [hh^\dagger]_{e\tau}$
 & $2v_\Delta^2 [hh^\dagger]_{\mu\tau}$\\
\hline\hline
 HI
 & $|\Delta m^2_{31}|
   \left(
    s_{13}^2 + \frac{\Delta m^2_{21}}{|\Delta m^2_{31}|} s_{12}^2
   \right)$
 & $\frac{|\Delta m^2_{31}|}{2}$
 & $\frac{|\Delta m^2_{31}|}{2}$
 & $\frac{|\Delta m^2_{31}|}{\sqrt{2}}
    \left(
     s_{13}
     + \frac{1}{\,2\,}
       \frac{\Delta m^2_{21}}{|\Delta m^2_{31}|} \sin{2\theta_{12}}
    \right)$
 & $\frac{|\Delta m^2_{31}|}{\sqrt{2}}
   \left(
     s_{13}
     - \frac{1}{\,2\,}
       \frac{\Delta m^2_{21}}{|\Delta m^2_{31}|} \sin{2\theta_{12}}
   \right)$
 & $\frac{|\Delta m^2_{31}|}{2}$\\
\hline
IN
 & $|\Delta m^2_{31}|$
 & $\frac{|\Delta m^2_{31}|}{2}$
 & $\frac{|\Delta m^2_{31}|}{2}$
 & $-\frac{|\Delta m^2_{31}|}{\sqrt{2}}
   \left(
     s_{13}
     - \frac{1}{\,2\,}
             \frac{\Delta m^2_{21}}{|\Delta m^2_{31}|} \sin{2\theta_{12}}
   \right)$
 & $-\frac{|\Delta m^2_{31}|}{\sqrt{2}}
   \left(
    s_{13}
     + \frac{1}{\,2\,}
             \frac{\Delta m^2_{21}}{|\Delta m^2_{31}|} \sin{2\theta_{12}}
   \right)$
 & $-\frac{|\Delta m^2_{31}|}{2}$\\
\hline
DG
 & $m_1^2$
 & $m_1^2$
 & $m_1^2$
 & $\frac{\Delta m^2_{31}}{\sqrt{2}}
   ( s_{13}
     + \frac{1}{\,2\,}
       \frac{\Delta m^2_{21}}{\Delta m^2_{31}} \sin{2\theta_{12}} )$
 & $\frac{\Delta m^2_{31}}{\sqrt{2}}
   ( s_{13}
     - \frac{1}{\,2\,}
       \frac{\Delta m^2_{21}}{\Delta m^2_{31}} \sin{2\theta_{12}} )$
 & $\frac{\Delta m^2_{31}}{2}$
\end{tabular}
\end{center}
\caption{Approximate forms of $2 v_\Delta (hh^\dagger)_{ij}$.}
\label{tab:hhij}
\end{sidewaystable}

\end{document}